\documentclass[
aps,
prapplied,
reprint,
amsmath,amssymb,
superscriptaddress
]{revtex4-2}

\usepackage{cancel}
\usepackage{graphicx}
\usepackage{dcolumn}
\usepackage{bm}

\usepackage{color,soul} 
\usepackage{comment} 	
\usepackage{xcolor} 

\usepackage{verbatim} 
\usepackage[bookmarks]{hyperref} 	

\setul{0.3ex}{0.25ex} 
\setulcolor{magenta}

\begin{document}
	
	
	\title{Universal length dependence of tensile stress in nanomechanical string resonators}
	
	\author{Maximilian B{\"u}ckle}
	\altaffiliation{Both authors contributed equally to this work.}
	\affiliation{%
		Department of Physics, University of Konstanz, 78457 Konstanz, Germany
	}%

	\author{Yannick S. Kla{\ss}}%
	\altaffiliation{Both authors contributed equally to this work.}
	\affiliation{%
		Department of Physics, University of Konstanz, 78457 Konstanz, Germany
	}%

	\author{Felix B. N{\"a}gele}%
	\affiliation{%
		Department of Physics, University of Konstanz, 78457 Konstanz, Germany
	}%
	
	\author{R\'{e}my Braive}
	\affiliation{%
		Centre de Nanosciences et de Nanotechnologies, CNRS, Universit\'{e} Paris-Sud, Universit\'{e} Paris-Saclay, 91767 Palaiseau, France
	}%
	\affiliation{%
		Universit\'{e} de Paris, 75207 Paris Cedex 13, France
	}%
	
	\author{Eva M. Weig}
	\email{eva.weig@tum.de}
	\affiliation{%
		Department of Physics, University of Konstanz, 78457 Konstanz, Germany
	}%
	\affiliation{%
		Department of Electrical and Computer Engineering, Technical University of Munich, 80333 Munich, Germany
	}%
	\affiliation{
		Munich Center for Quantum Science and Technology (MCQST), 80799 Munich, Germany
	}%

	\date{\today}
	
	\begin{abstract}
		
		We investigate the tensile stress in freely suspended nanomechanical string resonators, and observe a material-independent dependence on the resonator length. We compare strongly stressed string resonators fabricated from four different material systems based on amorphous silicon nitride, crystalline silicon carbide as well as crystalline indium gallium phosphide. The tensile stress is found to increase by approximately $50$\,\% for shorter resonators.
		We establish a simple elastic model to describe the observed length dependence of the tensile stress. The model accurately describes our experimental data.
		This opens a perspective for stress-engineering the mechanical quality factor of nanomechanical string resonators.

	\end{abstract}
	
	\maketitle
	
	
	%
	%
	
	\section{Introduction}
	
	Over the past decades, nanoelectromechanical systems (NEMS) have received considerable attention as versatile elements in mesoscopic devices. 
	For example, they are well-suited for sensor applications, and have been shown to act as highly sensitive mass 
	\cite{hanay2012single,chaste2012nanomechanical}, force \cite{mamin2001sub,Rugar2004,moser2013ultrasensitive,Simonsen2019}, or infrared \cite{piller2019nanoelectromechanical} detectors. 
	Even more, they serve as amplifiers or oscillators. 
	On the other hand, nanomechanical systems lend themselves to the exploration of fundamental physical phenomena both in the classical \cite{Yang2019,huber2020spectral,Karg2020} and quantum regimes
	\cite{sudhir2017quantum,rossi2018measurement,mason2019continuous,Guo2019}, with the prospect of serving as hybrid transducer
	\cite{Andrews2014} or as storage element \cite{OConnell2010} in future quantum technologies. 
	
	Many of the underlying devices are based on a one-dimensional string or two-dimensional membrane resonators  \cite{Thompson2008,Wilson2009,Joeckel2014,Ftouni2015,Fink2016,Serra2016,Norte2016,Reinhardt2016,sudhir2017quantum,Tsaturyan2017NatNano-UltracoherentNanomechSoftclamping,Ghadimi2018StrainEngineering,rossi2018measurement,mason2019continuous,Simonsen2019,Guo2019,Yang2019,yuksel2019nonlinear,huber2020spectral,Karg2020}
	that are both characterized by a strong intrinsic tensile stress in the device layer. Tensile stress in the device layer enables remarkably large mechanical quality factors as a result of a process commonly referred to as dissipation dilution \cite{GonzalezSaulson1994BrownianMotionAnelasticWire,Unterreithmeier2010PRL-DampingNanomechRes,Yu2012PRL-ControlMaterialDampingHighQMembraneMicrores,Villanueva2014PRL-EvidenceSurfaceLossSiN}. 
	This process relies on the stress-induced increase of the stored vibrational energy of the resonator, while the dissipated energy remains largely unaffected. 
	
	Several approaches have been pursued to suppress the limiting dissipation mechanism and to further enhance the potential of this class of nanomechanical resonators. For example, clamping losses can be reduced by means of mechanical impedance mismatch engineering \cite{bib:Rieger2014} 
	that is exploited, e.g., in trampoline resonators \cite{Kleckner2011,Reinhardt2016,Norte2016}. In addition, phononic bandgaps have been employed to reduce clamping losses \cite{Alegre2011,yu2014phononic,Tsaturyan2014}. Recently, soft clamping has been presented as an innovative approach of boosting the quality factor \cite{Tsaturyan2017NatNano-UltracoherentNanomechSoftclamping}. It is also based on phononic bandgap engineering, but additionally prevents mechanical strain at the interface between the resonator and its clamping points and thus enhances the dissipation dilution. 
	Soft-clamped membranes have recently enabled measurement-based quantum control \cite{rossi2018measurement} or measurements below the standard quantum limit \cite{mason2019continuous}, and may find application in magnetic resonance force imaging \cite{Simonsen2019}. 
	For string resonators, soft clamping can be beneficially combined with stress engineering \cite{Ghadimi2018StrainEngineering,fedorov2019generalized}, which boosts the tensile stress to close to the mechanical limits and enables quality factors $Q$ of around $800$ million and a $Q \times$ frequency product of more than $ 10^{15}\,$Hz. 
	These developments pave the way towards an ever increasing sensitivity and the observation of mechanical quantum phenomena at room temperature.
	
	Here we describe a previously unappreciated aspect of nanomechanical string resonators: the one-dimensional tensile stress in the string is not solely determined by elastic material properties, but significantly depends on its length as well as other geometric parameters. This allows one to increase the tensile stress by approximately $50$\,\% by using shorter strings and thus boost the dissipation dilution. The observed length dependence of the tensile stress is material independent. We demonstrate length-dependent tensile stress for nanomechanical resonators fabricated from four different wafers featuring the three complementary, tensile-stressed device layers silicon nitride (SiN), silicon carbide (SiC), and indium gallium phosphide (InGaP). A simple elastic model is developed that captures the observed features. It describes the geometric reconstruction of the string resonator by a combination of two effects that determine the stress distribution in the device layer: the vertical release of the string leads to a deformation of the clamping structure, while the subsequent lateral release undercuts the clamping pads. Our results entail important insights for strain engineering of nanomechanical devices and may be exploited to boost the mechanical quality factor of string resonators.

	\section{Experimental results}
	
	We investigate nanomechanical string resonators fabricated from three distinct, strongly stressed material platforms, namely amorphous silicon nitride (SiN), crystalline silicon carbide (SiC), and crystalline indium gallium phosphide (InGaP). The $100$\,nm thick amorphous stoichiometric SiN film is deposited by low-pressure chemical vapor deposition on top of two different substrates, a fused silica wafer (material denoted SiN-FS) and a sacrificial SiO$_2$ layer atop a silicon wafer (SiN-Si). The $110$\,nm thick film of crystalline 3C-SiC is epitaxially grown on a Si(111) wafer (SiC). The III-V heterostructure hosts a $100$\,nm thick In$_{0.415}$Ga$_{0.585}$P film epitaxially grown atop a sacrificial layer of Al$_y$Ga$_{1-y}$As on a GaAs wafer.
	See Tab.~S.1. of the Supplemental Material \cite{siCite} for the full details of all wafers.
	
\begin{figure}[t!]
		\includegraphics[width=0.9\linewidth]{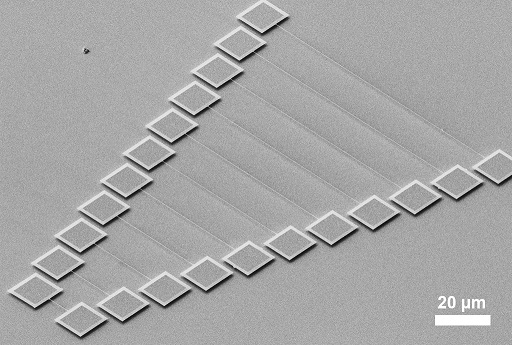}
		\caption{
			\label{fig:SEM} 
			Scanning electron micrograph of a series of nano\-string resonators with lengths increasing from $10$ to $110\,\mu$m in steps of $10\,\mu$m. 
		}
	\end{figure}

	Series of nanomechanical string resonators such as that depicted in Fig.~\ref{fig:SEM} are defined in all four material systems. The length of the strings increases from $10$ to $110\,\mu$m in steps of $10\,\mu$m. 
	The resonators are characterized using piezo actuation and interferometric detection. For all wafers, we record the resonance curves of the out-of-plane flexural modes in the linear response regime. The measurements are performed at room temperature and inside a vacuum chamber (pressure less than $10^{-3}$\,mbar) to avoid gas damping.
	
\begin{figure}[t!]
		\includegraphics[width=0.8\linewidth]{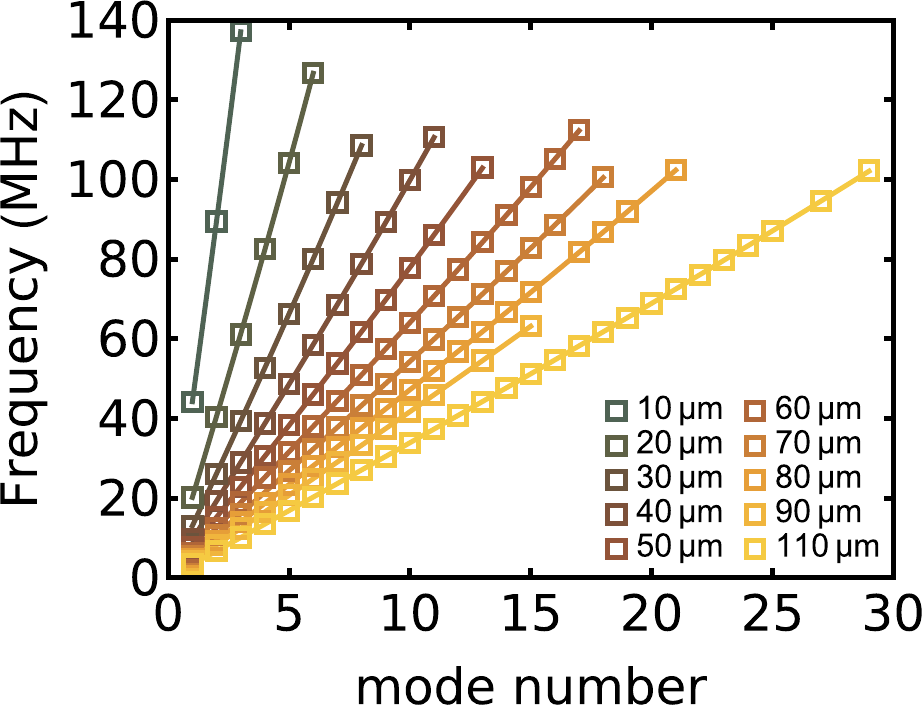}
		\caption{
			\label{fig:ModeFreq} 
			Eigenfrequencies of the out-of-plane modes as a function of the mode number for SiN string resonators on fused silica (SiN-FS). The resonator lengths range from $10$ to $110\,\mu$m.
			Fits of the eigenfrequencies using Eq.~(\ref{eq:frequency}) are included as solid lines.
		}
	\end{figure}

	For each resonator length we probe the fundamental out-of-plane mode as well as a series of up to $30$ higher-order modes, and determine the corresponding eigenfrequencies by Lorentzian fits, as shown for the case of SiN-FS in Fig.~\ref{fig:ModeFreq}.
	Following Euler-Bernoulli beam theory of a doubly clamped string with simply supported boundary conditions \cite{Timoshenko1990-VibrationProblemsEngineering,Cleland2002foundations}, we can conveniently express the eigenfrequency of the $n$-th mode as
	\begin{equation}
		f_n = \frac{n^2 \pi}{2 L^2} \sqrt{\frac{E_1 h_1^2}{12\rho}}\sqrt{1+ \frac{12 \sigma L^2}{n^2 \pi^2 E_1 h_1^2}} \label{eq:frequency}
	\end{equation}
	where $E_1$ is Young's modulus, $\rho$ is the mass density, $h_1$ is the thickness of the resonator, and $\sigma$ is the tensile stress of the string.
	
\begin{figure}[t!]
		\includegraphics[width=0.8\linewidth]{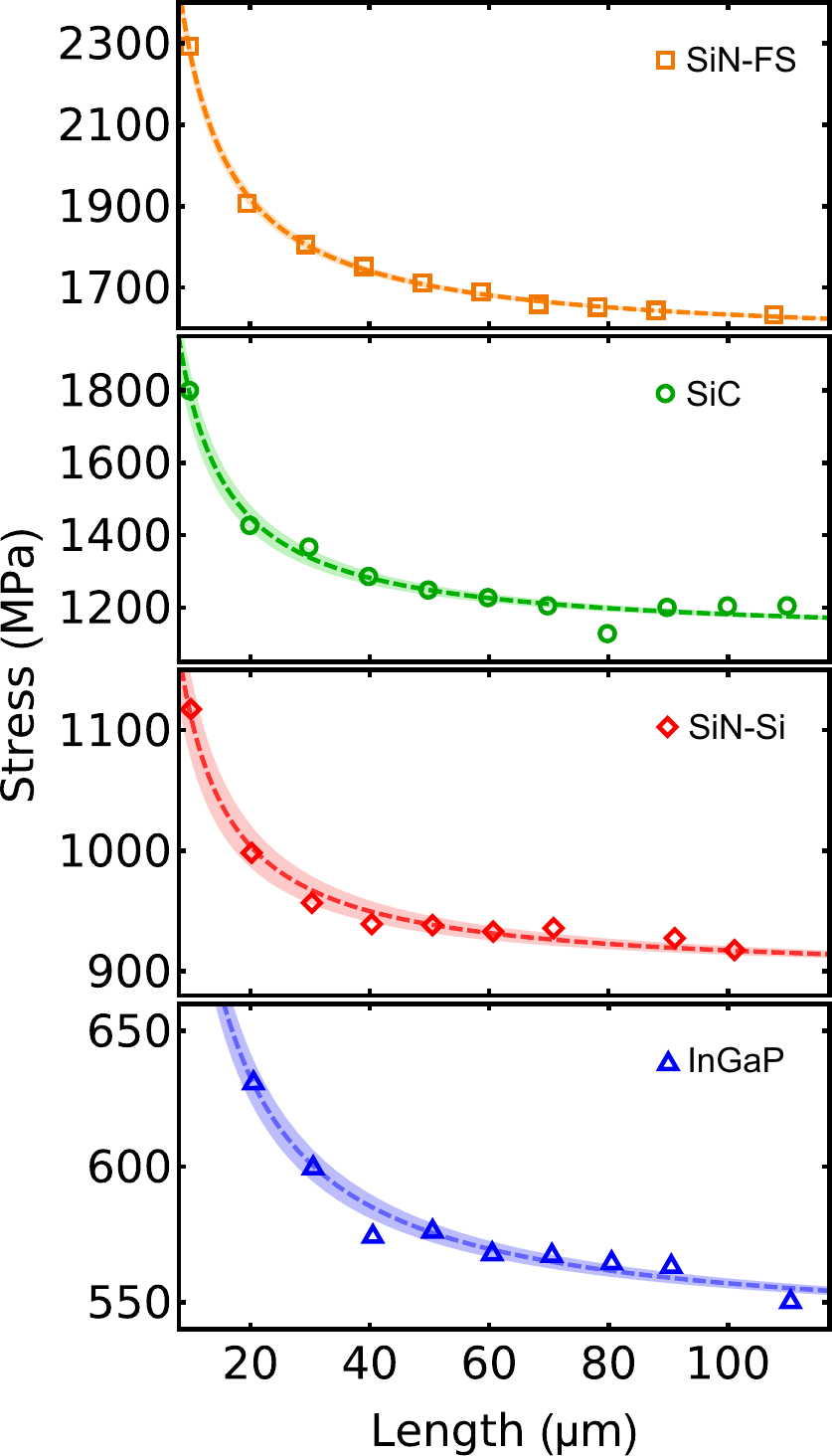}
		\caption{
			\label{fig:sigL} 
			Experimentally determined tensile stress as a function of the length of the nanostring for all four material systems. 
			Fits of Eq.~(\ref{eq:sigLM}) are included as solid lines. The obtained fit parameters are summarized in Tab.~\ref{tab:disp}.
			The shaded areas indicate the uncertainty resulting from measurement errors of the pedestal height $h_0$ and undercut $a_\mathrm{uc}$.
		}
	\end{figure}

	The resonance frequencies shown in Fig.\,\ref{fig:ModeFreq} are fitted with Eq.~(\ref{eq:frequency}). To a very good approximation, the eigenfrequencies of all resonators scale linearly with the mode number as expected for stress-dominated nanostrings for which  
	$f_n \approx (n/2 L) \sqrt{\sigma / \rho}$.
	The fits of the eigenfrequencies as a function of mode number allow to extract the tensile stress of each nanostring resonator. The obtained values are shown as a function of the resonator length for all four materials in Fig.\,\ref{fig:sigL}. Clearly, the tensile stress is not constant, but decreases for increasing resonator length. The same qualitative behavior is observed in all four material systems.
	
	\section{Elastic model} 
	
\begin{figure}[ht]
		\includegraphics[width=\linewidth]{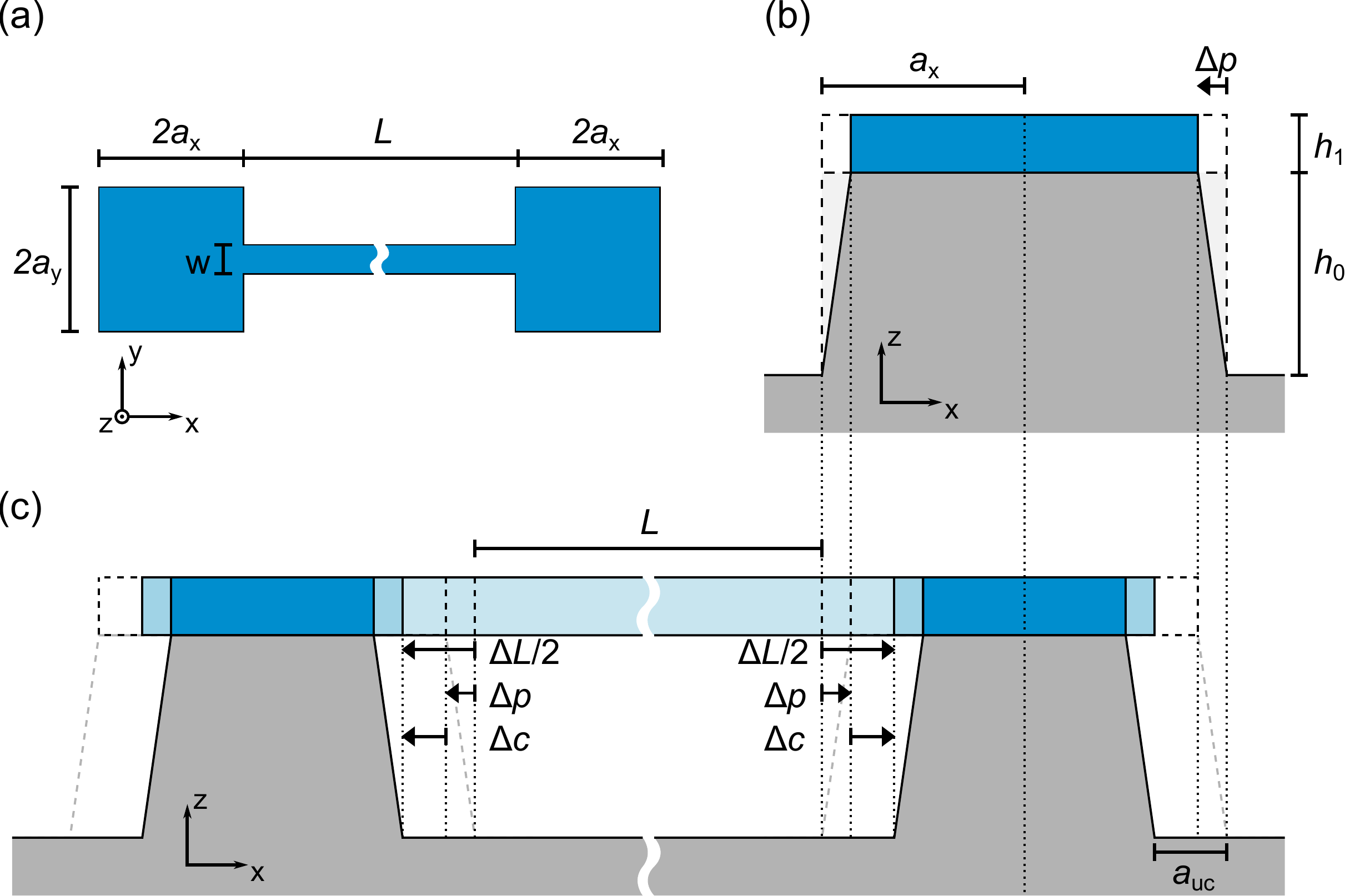}
		\caption{
			\label{fig:toymodel} 
			Sample geometry and parameters of the model. 
			(a) Lithographic dimensions of the nanostring and its clamping pads.  
			(b) Cross section through the clamping structure illustrating the shearing contraction of the pedestal following the vertical release of the structure.
			(c) Cross section through the clamping structure illustrating the lateral contraction of the undercut areas of the clamping pad following the horizontal release. Combining the vertical and horizontal releases leads to the string's length change of $ \Delta L $. Areas supported by a pedestal are colored in dark blue, undercut areas of the clamping pads are indicated by a lighter color and the string is marked with the lightest blue. Dotted lines serve as guides to the eye.
		}
	\end{figure}

	A length dependence of the tensile stress in a nanostring has not been reported in the literature. We have developed a simple theoretical model to quantify this previously unappreciated phenomenon. Our model is based on elastic theory. As such, it is material independent and can be applied to all material systems under investigation. The model assumes a prismatic string of length $L$, width $w$ and thickness $h_1$. Its cross-sectional area is $A_{\textrm{s}} = w \, h_1$. On both ends, the string is attached to a rectangular clamping structure. It consists of a clamping pad in the device layer with lithographic dimensions $2 a_x$ and $2 a_y$, as well as thickness $h_1$ (Fig.~\ref{fig:toymodel}(a)), which is supported by a pedestal of height $h_0$ in the underlying sacrificial or substrate layer (Fig.~\ref{fig:toymodel}(b) and (c)). As a result of the wet etching process required to suspend the nanostrings, the clamping pads exhibit a certain undercut $a_\mathrm{uc}$, i.e. the width of the pad $2a_x$ (in $x$ direction, the width $a_y$ in $y$ direction may differ) is larger than that of the remaining pedestal $2 a_\mathrm{p} = 2a_x - 2a_\mathrm{uc}$. The cross-sectional area of the clamping pad (in $yz$-plane) is  $A_{\textrm{p}} = 2a_y \, h_1$. The geometric parameters of the four investigated samples are summarized in Tab.~\ref{tab:ModelParameters}. The thickness of the device layer $h_1$ is obtained from the wafer growth protocol, whereas the height of the pedestal $h_0$ is determined by atomic force microscopy. The half-widths of the clamping pad $a_x$ and $a_y$, the undercut $ a_\mathrm{uc} $ and the width of the nanostring $w$ have been extracted using electron beam microscopy. The elastic and material parameters of the samples \cite{DosterUnpub,maluf2002introduction,li1987single,henisch2013silicon,Ioffe1999ShurEtAl-HandbookSeriesSemiconductorParametersVOL2} are listed in Tab.~S.2. in the Supplemental Material \cite{siCite}.
	
	\begin{table}
		\caption{
			\label{tab:ModelParameters}
			Geometric parameters of the investigated samples.
		}
		\begin{ruledtabular}
			\begin{tabular}{crrrrrr}
				& \multicolumn{1}{c}{\textrm{$h_1$}} & \multicolumn{1}{c}{\textrm{$h_0$}} &  
				\multicolumn{1}{c}{\textrm{$2a_x$}} & 
				\multicolumn{1}{c}{\textrm{$2a_y$}} &
				\multicolumn{1}{c}{\textrm{$a_\mathrm{uc}$}} &
				\multicolumn{1}{c}{\textrm{$w$}} \\
				& \multicolumn{1}{c}{\textrm{(nm)}} & \multicolumn{1}{c}{\textrm{(nm)}} & \multicolumn{1}{c}{\textrm{($ \mu $m)}} & \multicolumn{1}{c}{\textrm{($ \mu $m)}} & \multicolumn{1}{c}{\textrm{(nm)}} & \multicolumn{1}{c}{\textrm{(nm)}}  \\
				\hline
				SiN-FS	& 100(5) & 460(20)	& 13.7(2) & 13.6(2)	& 570(100) 	& 420(25) \\
				SiN-Si	& 100(5) & 365(20)	& 14.1(2) & 15.0(2) & 410(150) 	& 340(25) \\
				SiC 	& 110(15)& 570(40) 	& 14.2(2) & 15.0(2) & 860(150) 	& 360(30) \\
				InGaP 	& 100(1) & 990(10)	& 12.7(2) & 13.3(2) & 640(170)	& 250(15) \\
			\end{tabular}
		\end{ruledtabular}
	\end{table}
	
	As we show in the following, the tensile stress in the device layer atop an unstressed sacrificial layer or substrate gives rise to a balance of forces that in turn leads to a length- and geometry-dependent change in the one-dimensional tensile stress of the nanostring. To quantify the contributing forces, we roughly follow the process sequence required to fabricate a freely suspended nanostring. 
	First, we consider the vertical release of the nanostructure. It comprises all vertical etching contributions, notably the anisotropic dry etch employed for the pattern transfer following electron beam lithography. Note that the subsequent wet etching process releasing the nanostring may also have a vertical etching component, e.g., for the case of an isotropic wet etch. The vertical release penetrates both the device layer and the sacrificial or substrate layer and defines the height of the pedestal $h_0$.
	Following this vertical release, the tensile-stressed device layer will contract and induce a certain amount of shear in the pedestal (Fig.~\ref{fig:toymodel}(b)). As a result, the tensile stress in the pad relaxes to a value $\sigma_\mathrm{p}$. 
	
	Second, the lateral release is considered. It accounts for the lateral etching during the nanostring release.
	As a result of this lateral release, the two-dimensional stress in the nanostring relaxes in the direction perpendicular to the string. At the same time, the undercut parts of the tensile-stressed clamping pad contract, which applies additional stress on the nanostring (Fig.~\ref{fig:toymodel}(c)). The combination of the described effects gives rise to the tensile stress experienced by the nanostring $\sigma$ (see Eq.~(\ref{eq:frequency}) and Fig.~\ref{fig:sigL}). The model assumes a clear separation between the vertical and lateral release, which will be described in Secs.~\ref{Pedestal} and~\ref{Undercut}, respectively, and neglects geometric and elastic reconfigurations of the sheared pedestal and stressed clamping pad arising from the lateral releases, which we can safely assume to be small.
	
	\subsection{Pedestal shear from vertical release}\label{Pedestal}
	
	To evaluate the shear of the pedestal induced by the vertical release of the structure, we first consider an isolated clamping structure and focus on its cross section along the $x$-$z$ direction as indicated in Fig.~\ref{fig:toymodel}(b). 
	The resonator will be included at a later stage. 
	Following the vertical release of the structure, the strong tensile stress in the device layer leads to a contraction of the clamping pad in order to minimize internal forces. 
	This contraction produces an increasing shear of the pedestal. 
	The reconfiguration of the clamping structure stops once equilibrium between the reduced tensile force and the counteracting shearing force is reached. 
	The shear stress $\tau$ of such a shear-constrained material system can be expressed as \cite{Taylor1962} 
	
	\begin{equation}
		\tau = \sigma_{\mathrm{2D}} h_1  k  \tanh\left(k a_{x}\right) , \quad k = \sqrt{\frac{G_0}{h_0}  \frac{1}{E_1 h_1}} \label{eq:mSL:tau}
	\end{equation}
	where $h_0$ and $h_1$ are the heights of the pedestal and the clamping pad, respectively, $G_0$ is the shear modulus of the pedestal, $E_1$ is Young's modulus of the clamping pad, and $\sigma_{2D}$ is the initial two-dimensional stress in the device layer.
	This results in the maximum contraction of the clamping pad $\Delta p$ from its original half-width $a_\mathrm{x}$:  
	\begin{equation}
		\Delta p = \frac{h_0}{G_0} \tau = \frac{\sigma_{\mathrm{2D}}}{E_1 k} \tanh\left(k a_{x}\right) . 
		\label{umax}
	\end{equation}
	In consequence, the tensile stress in the clamping pad is reduced to $\sigma_\mathrm{p} = \sigma_{2D} - E_1 \frac{\Delta p}{a_{x}} $ according to Hooke's law. Note that a similar model that also accounts for additional shear in the device layer is presented in Ref.~\cite{suhir2008interfacial}.

	For the sake of simplicity, we neglect the minute counterforce exerted by the presence of the resonator, which will lead to a slightly reduced contraction of the pad to which it is attached. Similarly, the effect of shear of the pedestal underneath the resonator will be ignored. 
	
	An experimental verification of the contraction of the clamping pad following the vertical release is discussed in the Supplemental Material \cite{siCite}. 
	
	\subsection{Undercut of clamping pads from lateral release}\label{Undercut}
	
	The lateral release of the nanostrings results in an undercut of the clamping pads. More specifically, the width of the pedestal is reduced by $a_\mathrm{uc}$ from either side such that the rim of the clamping pad gets freely suspended as shown in Fig.~\ref{fig:toymodel}(c). This enables a relaxation of the tensile force in the undercut parts of the pads that gives rise to a contraction by an amount $ \Delta c $. The resulting contracting force acting on the interface between the clamping pad and the nanostring can be expressed as
	\begin{equation}
		F_\mathrm{c} = \sigma_\mathrm{p} A_\mathrm{p} - E_1 \frac{\Delta c}{a_\mathrm{uc}} A_\mathrm{p},
	\end{equation}
	where $\sigma_\mathrm{p}$ is the remaining tensile stress in the clamping pad following the vertical release, and $E_1 \frac{\Delta c}{a_\mathrm{uc}}$ is its reduction in the undercut part of the clamping pad, again according to Hooke's law. 
	Note that in the absence of the nanostring, the suspended part of the clamping pad fully relaxes such that $F_\mathrm{c} = 0$.   
	In the presence of the nanostring, however, the contracting force of the clamping pad is counteracted by a second force acting on the interface between the clamping pad and the nanostring which is associated with the elongation $\Delta L$ of the nanostring

	\begin{equation}
		F_\mathrm{s} = \sigma_\infty A_\mathrm{s} + E_1 \frac{\Delta L}{L} A_\mathrm{s},
	\end{equation}
	where $\sigma_\infty$ is the one-dimensional stress of an infinitely long nanostring after the lateral release, and $E_1 \frac{\Delta L}{L} $ is its modification according to Hooke's law. 
	
	The equilibrium condition for the clamping pad - nanostring interface 
	\begin{equation}
		F_c = F_s  \label{eq:interfaceCondition}
	\end{equation}
	determines the final geometric reconfiguration of the clamping pad and the string, under the boundary condition that the total length of the compound between the centers of the clamping pads has to be conserved,
	\begin{equation}
		2 \Delta p + 2 \Delta c = \Delta L . \label{eq:conservLength}
	\end{equation}

	Equations (\ref{eq:interfaceCondition}) and (\ref{eq:conservLength}) form a second-order system of linear equations with the unknown parameters $ \Delta L $ and $ \Delta c $. The third unknown $\Delta p$ is determined using Eq.~(\ref{umax}). Solving for the elongation of the resonator yields
	\begin{equation}
		\Delta L = 2 L \, \frac{\left( A_\mathrm{p} a_\mathrm{uc} \sigma_\mathrm{p} + A_\mathrm{p} E_1 \Delta p - A_\mathrm{s} a_\mathrm{uc} \sigma_\infty \right)}{E_1 \left( 2 A_\mathrm{s} a_\mathrm{uc} + A_\mathrm{p} L \right)} .
		\label{eq:delL}
	\end{equation}

	This length change of the resonator directly translates into an additional strain $ \varepsilon=\Delta L/L $, giving rise to a length-dependent stress $ \sigma(L) $ of the doubly clamped string resonator via Hooke's law
	\begin{equation}
		\sigma(L) = \sigma_\infty + E_1 \frac{\Delta L}{L}. \label{eq:sigLM} 
	\end{equation}

	\section{Discussion}
	
	To validate the theoretical model, we fit Eq.~(\ref{eq:sigLM}) to the experimental data measured on all four material systems, using the geometric and material parameters specified in Tabs.~\ref{tab:ModelParameters} and S.2 of the Supplemental Material \cite{siCite}. The initial two-dimensional stress 
	$\sigma_\mathrm{2D}$ can be calculated from the epitaxial lattice mismatch of the crystalline InGaP sample. The mismatch of the lattice constants of the In$_{1-x}$Ga$_x$P layer with respect to the underlying Al$_{y}$Ga$_{1-y}$As sacrificial layer induces an in-plane strain $\varepsilon^\parallel(x) = (a^\parallel_\mathrm{L} - a^\infty_\mathrm{L}(x) )/a^\infty_\mathrm{L}(x)$, where $a^\infty_\mathrm{L}(x)$ is the lattice constant of In$_{1-x}$Ga$_x$P and $a^\parallel_\mathrm{L}$ is the in-plane lattice constant of the strained In$_{1-x}$Ga$_x$P film. This allows to compute the two-dimensional stress of the thin In$_{1-x}$Ga$_x$P layer $\sigma_\mathrm{2D} = \varepsilon^\parallel E_1 /(1-\nu_1)$ \cite{Bueckle2018APL-StressControl}. The ratio $ E_1/(1-\nu_1) $ including the Poisson ratio $\nu_1$ (see Tab.~S.2. in the Supplemental Material \cite{siCite}) represents the biaxial modulus of the stressed thin film, which is required as no stress occurs in the $z$ direction normal to the substrate. The obtained $\sigma_\mathrm{2D}$ value of $0.95$\,GPa has also been used as an input parameter for the model.  
	In principle, the same argument can be made for SiC which is also an epitaxially grown crystalline thin-film material. However, the crystallization of 3C-SiC atop a Si wafer is more complex than that of the III-V heterostructures. SiC and Si feature the same crystal structure, but exhibit a lattice mismatch of approximately $20\,\%$. This implies a nontrivial, commensurate growth of the SiC film that strongly depends on growth conditions, such that the two-dimensional stress cannot be predicted from the crystal structure \cite{pakula2004fabrication,zorman2008micro,iacopi2013evidence,Iacopi2013APL-QrientationDependentStressRelax3C-SiCFilms,Kermany2016JAP-FactorsAffectionFQproduct3CSiC,Romero2020}. 
	Hence, we set $ \sigma_{\mathrm{2D}} $ as an additional fit parameter for SiC. 
	The same applies for the amorphous thin-film materials SiN-FS and SiN-Si.
	The one-dimensional stress  $ \sigma_\infty $ is employed as fit parameter for all material systems. 
	
	The results of the fits are included in Fig.~\ref{fig:sigL} as solid lines. The shaded area represents the model's uncertainty arising from the error of the input parameters.
	As long as $A_\textrm{s} \ll A_\textrm{p}$ and $a_\textrm{uc} \ll L$, the length dependence of Eq.~(\ref{eq:sigLM}) can be approximated as $ \sigma(L) \propto 1/L $. This holds true for all nanostrings under investigation, such that a $1/L$ dependence of the stress can be assumed.  
	We find remarkable agreement between the model and the experimental data. This is particularly noteworthy for the case of the InGaP samples for which only one fit parameter, $ \sigma_\infty $, is employed, which, in the above approximation of small $A_\textrm{s} a_\textrm{uc}$ corresponds to a vertical offset and thus the limit $\sigma(L\rightarrow \infty)$. 
	Also the results for SiN and SiC, which involve two fitting parameters, show good agreement between the model and the experimental data. Again, $ \sigma_\infty $ can be interpreted as the tensile stress of an infinitely long string, whereas the two-dimensional stress in the as-grown device layer $\sigma_\mathrm{2D}$ can, at least to some extent, be compared to literature values.
	
	In Table~\ref{tab:disp} we summarize the parameters obtained from the elastic model as well as the fit parameters for the case of the longest strings. 
	The as-grown two-dimensional stress in low-pressure chemical vapor deposition (LPCVD) grown stoichiometric SiN on silicon is found to depend on growth conditions, but has been reported to amount to $1.1$\,GPa~\cite{Ghadimi2018StrainEngineering,Bereyhi2019} and $1.4$\,GPa~\cite{Unterrreithmeier2009Nature-TransductionDielectricForces}, which is close to the value found here. 
	The same applies for high stress 3C-SiC(111), for which an as-grown two-dimensional stress of $1.3$\,GPa has been reported~\cite{iacopi2013evidence}, which is somewhat lower than our result. The growth of high stress SiN on a fused silica substrate is poorly characterized, and no comparison with the literature could be obtained. 
	Certainly, all observed two-dimensional stress values 
	are well within the yield strength of the respective material, which amounts to approximately $6-7$\,GPa (or even $12$\,GPa according to Ref.~\cite{Yang2002}) for high stress LPCVD-deposited SiN \cite{Kaushik2005,Ghadimi2018StrainEngineering,Bereyhi2019}, and $21$\,GPa for SiC \cite{Petersen1982}.

	
	\begin{table}
		\caption{
			\label{tab:disp}
			Parameters of the elastic model for the case of a long string, as well as two-dimensional and one-dimensional stress.  
		}
		\begin{ruledtabular}
			\begin{tabular}{cddddd}
				& \multicolumn{1}{c}{\textrm{$\Delta p$}} & \multicolumn{1}{c}{\textrm{$\Delta c$}} & \multicolumn{1}{c}{\textrm{$\Delta L$}} & 
				\multicolumn{1}{c}{\textrm{$\sigma_\mathrm{2D}$}} & \multicolumn{1}{c}{\textrm{$\sigma_\infty$}} \\
				& \multicolumn{1}{c}{\textrm{(nm)}} & \multicolumn{1}{c}{\textrm{(nm)}} & \multicolumn{1}{c}{\textrm{(nm)}} & \multicolumn{1}{c}{\textrm{(GPa)}} & \multicolumn{1}{c}{\textrm{(MPa)}} \\
				\hline
				SiN-FS 	& 8 	& 6  	& 27 	& 3.1\footnotemark[1] & 1560\footnotemark[1] \\
				SiN-Si 	& 3 	& 2 & 8 & 1.2\footnotemark[1] & 900\footnotemark[1]  \\
				SiC  	& 3 	& 5 & 16	& 2.5\footnotemark[1] & 1110\footnotemark[1] \\
				InGaP 	& 3 	& 5 	& 15 	& 0.95\footnotemark[2] & 540\footnotemark[1] \\
			\end{tabular}
		\end{ruledtabular}
		\footnotemark[1]{From fit}
		\footnotemark[2]{Calculated with $\sigma_\mathrm{2D}=\varepsilon^\parallel E_1/(1-\nu_1)$ \cite{Bueckle2018APL-StressControl}}
	\end{table}
	
	A more general consideration of the length dependence of the tensile stress according to Eqs.~(\ref{eq:delL}) and (\ref{eq:sigLM}) reveals that two geometric parameters, the height of the pedestal $h_0$ and the undercut of the pedestal $a_\textrm{uc}$, dominate the stress enhancement of short nanostrings. This suggests that maximum tensile stress can be achieved for short strings with large $h_0$ and $a_\textrm{uc}$. However, it has to be noted that this limit can only be achieved for sufficiently large clamping pads avoiding a softening of the entire clamping structure under overly large undercuts, an unwanted side effect that is not accounted for in our model.
	
	Finally, we discuss the relation between $ \sigma_\infty $ and $ \sigma_{\mathrm{2D}} $. For a one-dimensional nanostring processed from a thin film under biaxial and isotropic stress, the one-dimensional stress follows from the initial two-dimensional stress according to  
	\begin{equation}
		\sigma_{\mathrm{1D}} = \sigma_{\mathrm{2D}} (1-\nu_1) \label{eq:sigma1D-2D}.
	\end{equation} 
	For the nanostrings under investigation, this simple picture does not hold, as the stress relaxation along the $y$ direction upon releasing the string assumes a more complicated stress configuration following the contraction of the device layer described in Sec.~\ref{Pedestal}. Not only does the contraction of the clamps by an amount $\Delta p$ reduced the two-dimensional stress in the clamping pads from $\sigma_{\mathrm{2D}}$ to $\sigma_{\mathrm{p}}$. A similar contraction also occurs along the $y$ direction of the string, such that the tensile stress in the string before the lateral release cannot be considered isotropic. 
	
	Additionally, we wish to note that high-resolution x-ray diffraction measurements performed on In$_{1-x}$Ga$_x$P wafers have shown a compositional variation in the direction normal to the substrate \cite{Bueckle2018APL-StressControl}. This can furthermore lead to strain gradients inside the device layer. A similar observation has been made for 3C-SiC in Ref.~\cite{Romero2020}. This suggests that a more thorough analysis of the length-dependent stress should assume a three-dimensional strain tensor accounting for a vertical strain gradient rather than a biaxial isotropic thin-film stress $\sigma_{\mathrm{2D}}$.

	\section{Conclusion}
	
	In conclusion, we report on the observation of a length dependence of the tensile stress in nanomechanical string resonators. This previously unappreciated effect is material independent, and experimentally observed on samples fabricated from four different wafers, featuring the three complementary material platforms amorphous silicon nitride, crystalline silicon carbide and crystalline indium gallium phosphide. We develop a simple elastic model that describes the observed $1/L$ dependence of the tensile stress, and which allows us to explain the observed length dependence by a combination of the elastic reconfiguration of the device under the vertical and lateral releases of the one-dimensional nanostring. The one-dimensional tensile stress relaxes to a value considerably smaller than the initial two-dimensional stress value for long strings. For shorter strings, this value increases by approximately $50$\,\%. Besides the length, particularly the height of the supporting pedestal and the size of the undercut of the clamping pads influence the resulting stress. Thus, the geometric parameters of the nanostring allow us to control the tensile stress, and enable stress engineering of the quality factor of the device without the need for complex phononic metamaterial processing.
	
	Data and analysis code are available onlschneg \cite{zenodo}.
	
	\begin{acknowledgments}
		We thank Ralf Messmer for technical support with the AFM measurements.
		Financial support from the European
		Unions Horizon 2020 programme for Research and Innovation
		under Grant Agreement No. 732894 (FET Proactive HOT) and the German Federal Ministry of Education
		and Research through Contract No. 13N14777 funded within
		the European QuantERA cofund project QuaSeRT is gratefully
		acknowledged. 
		We further acknowledge support from the Deutsche Forschungsgemeinschaft via
		the Collaborative Research Center SFB 1432 and via project WE 4721/1-1, as well as project QT-6 SPOC of the Baden-W\"urttemberg Foundation
		
		M.B. and Y.S.K. contributed equally to this work.
		
	\end{acknowledgments}

	\bibliography{main}

\begin{thebibliography}{60}%
\makeatletter
\providecommand \@ifxundefined [1]{%
 \@ifx{#1\undefined}
}%
\providecommand \@ifnum [1]{%
 \ifnum #1\expandafter \@firstoftwo
 \else \expandafter \@secondoftwo
 \fi
}%
\providecommand \@ifx [1]{%
 \ifx #1\expandafter \@firstoftwo
 \else \expandafter \@secondoftwo
 \fi
}%
\providecommand \natexlab [1]{#1}%
\providecommand \enquote  [1]{``#1''}%
\providecommand \bibnamefont  [1]{#1}%
\providecommand \bibfnamefont [1]{#1}%
\providecommand \citenamefont [1]{#1}%
\providecommand \href@noop [0]{\@secondoftwo}%
\providecommand \href [0]{\begingroup \@sanitize@url \@href}%
\providecommand \@href[1]{\@@startlink{#1}\@@href}%
\providecommand \@@href[1]{\endgroup#1\@@endlink}%
\providecommand \@sanitize@url [0]{\catcode `\\12\catcode `\$12\catcode
  `\&12\catcode `\#12\catcode `\^12\catcode `\_12\catcode `\%12\relax}%
\providecommand \@@startlink[1]{}%
\providecommand \@@endlink[0]{}%
\providecommand \url  [0]{\begingroup\@sanitize@url \@url }%
\providecommand \@url [1]{\endgroup\@href {#1}{\urlprefix }}%
\providecommand \urlprefix  [0]{URL }%
\providecommand \Eprint [0]{\href }%
\providecommand \doibase [0]{https://doi.org/}%
\providecommand \selectlanguage [0]{\@gobble}%
\providecommand \bibinfo  [0]{\@secondoftwo}%
\providecommand \bibfield  [0]{\@secondoftwo}%
\providecommand \translation [1]{[#1]}%
\providecommand \BibitemOpen [0]{}%
\providecommand \bibitemStop [0]{}%
\providecommand \bibitemNoStop [0]{.\EOS\space}%
\providecommand \EOS [0]{\spacefactor3000\relax}%
\providecommand \BibitemShut  [1]{\csname bibitem#1\endcsname}%
\let\auto@bib@innerbib\@empty
\bibitem [{\citenamefont {Hanay}\ \emph {et~al.}(2012)\citenamefont {Hanay},
  \citenamefont {Kelber}, \citenamefont {Naik}, \citenamefont {Chi},
  \citenamefont {Hentz}, \citenamefont {Bullard}, \citenamefont {Colinet},
  \citenamefont {Duraffourg},\ and\ \citenamefont {Roukes}}]{hanay2012single}%
  \BibitemOpen
  \bibfield  {author} {\bibinfo {author} {\bibfnamefont {M.~S.}\ \bibnamefont
  {Hanay}}, \bibinfo {author} {\bibfnamefont {S.}~\bibnamefont {Kelber}},
  \bibinfo {author} {\bibfnamefont {A.}~\bibnamefont {Naik}}, \bibinfo {author}
  {\bibfnamefont {D.}~\bibnamefont {Chi}}, \bibinfo {author} {\bibfnamefont
  {S.}~\bibnamefont {Hentz}}, \bibinfo {author} {\bibfnamefont
  {E.}~\bibnamefont {Bullard}}, \bibinfo {author} {\bibfnamefont
  {E.}~\bibnamefont {Colinet}}, \bibinfo {author} {\bibfnamefont
  {L.}~\bibnamefont {Duraffourg}},\ and\ \bibinfo {author} {\bibfnamefont
  {M.}~\bibnamefont {Roukes}},\ }\bibfield  {title} {\bibinfo {title}
  {Single-protein nanomechanical mass spectrometry in real time},\ }\href
  {https://doi.org/https://doi.org/10.1038/nnano.2012.119} {\bibfield
  {journal} {\bibinfo  {journal} {Nature nanotechnology}\ }\textbf {\bibinfo
  {volume} {7}},\ \bibinfo {pages} {602} (\bibinfo {year} {2012})}\BibitemShut
  {NoStop}%
\bibitem [{\citenamefont {Chaste}\ \emph {et~al.}(2012)\citenamefont {Chaste},
  \citenamefont {Eichler}, \citenamefont {Moser}, \citenamefont {Ceballos},
  \citenamefont {Rurali},\ and\ \citenamefont
  {Bachtold}}]{chaste2012nanomechanical}%
  \BibitemOpen
  \bibfield  {author} {\bibinfo {author} {\bibfnamefont {J.}~\bibnamefont
  {Chaste}}, \bibinfo {author} {\bibfnamefont {A.}~\bibnamefont {Eichler}},
  \bibinfo {author} {\bibfnamefont {J.}~\bibnamefont {Moser}}, \bibinfo
  {author} {\bibfnamefont {G.}~\bibnamefont {Ceballos}}, \bibinfo {author}
  {\bibfnamefont {R.}~\bibnamefont {Rurali}},\ and\ \bibinfo {author}
  {\bibfnamefont {A.}~\bibnamefont {Bachtold}},\ }\bibfield  {title} {\bibinfo
  {title} {A nanomechanical mass sensor with yoctogram resolution},\ }\href
  {https://doi.org/10.1038/nnano.2012.42} {\bibfield  {journal} {\bibinfo
  {journal} {Nature nanotechnology}\ }\textbf {\bibinfo {volume} {7}},\
  \bibinfo {pages} {301} (\bibinfo {year} {2012})}\BibitemShut {NoStop}%
\bibitem [{\citenamefont {Mamin}\ and\ \citenamefont
  {Rugar}(2001)}]{mamin2001sub}%
  \BibitemOpen
  \bibfield  {author} {\bibinfo {author} {\bibfnamefont {H.}~\bibnamefont
  {Mamin}}\ and\ \bibinfo {author} {\bibfnamefont {D.}~\bibnamefont {Rugar}},\
  }\bibfield  {title} {\bibinfo {title} {Sub-attonewton force detection at
  millikelvin temperatures},\ }\href
  {https://doi.org/https://doi.org/10.1063/1.1418256} {\bibfield  {journal}
  {\bibinfo  {journal} {Applied Physics Letters}\ }\textbf {\bibinfo {volume}
  {79}},\ \bibinfo {pages} {3358} (\bibinfo {year} {2001})}\BibitemShut
  {NoStop}%
\bibitem [{\citenamefont {Rugar}\ \emph {et~al.}(2004)\citenamefont {Rugar},
  \citenamefont {Budakian}, \citenamefont {Mamin},\ and\ \citenamefont
  {Chui}}]{Rugar2004}%
  \BibitemOpen
  \bibfield  {author} {\bibinfo {author} {\bibfnamefont {D.}~\bibnamefont
  {Rugar}}, \bibinfo {author} {\bibfnamefont {R.}~\bibnamefont {Budakian}},
  \bibinfo {author} {\bibfnamefont {H.~J.}\ \bibnamefont {Mamin}},\ and\
  \bibinfo {author} {\bibfnamefont {B.~W.}\ \bibnamefont {Chui}},\ }\bibfield
  {title} {\bibinfo {title} {Single spin detection by magnetic resonance force
  microscopy},\ }\href {https://doi.org/10.1038/nature02658} {\bibfield
  {journal} {\bibinfo  {journal} {Nature}\ }\textbf {\bibinfo {volume} {430}},\
  \bibinfo {pages} {329} (\bibinfo {year} {2004})}\BibitemShut {NoStop}%
\bibitem [{\citenamefont {Moser}\ \emph {et~al.}(2013)\citenamefont {Moser},
  \citenamefont {G{\"u}ttinger}, \citenamefont {Eichler}, \citenamefont
  {Esplandiu}, \citenamefont {Liu}, \citenamefont {Dykman},\ and\ \citenamefont
  {Bachtold}}]{moser2013ultrasensitive}%
  \BibitemOpen
  \bibfield  {author} {\bibinfo {author} {\bibfnamefont {J.}~\bibnamefont
  {Moser}}, \bibinfo {author} {\bibfnamefont {J.}~\bibnamefont
  {G{\"u}ttinger}}, \bibinfo {author} {\bibfnamefont {A.}~\bibnamefont
  {Eichler}}, \bibinfo {author} {\bibfnamefont {M.~J.}\ \bibnamefont
  {Esplandiu}}, \bibinfo {author} {\bibfnamefont {D.}~\bibnamefont {Liu}},
  \bibinfo {author} {\bibfnamefont {M.}~\bibnamefont {Dykman}},\ and\ \bibinfo
  {author} {\bibfnamefont {A.}~\bibnamefont {Bachtold}},\ }\bibfield  {title}
  {\bibinfo {title} {Ultrasensitive force detection with a nanotube mechanical
  resonator},\ }\href {https://doi.org/10.1038/nnano.2013.97} {\bibfield
  {journal} {\bibinfo  {journal} {Nature nanotechnology}\ }\textbf {\bibinfo
  {volume} {8}},\ \bibinfo {pages} {493} (\bibinfo {year} {2013})}\BibitemShut
  {NoStop}%
\bibitem [{\citenamefont {Simonsen}\ \emph {et~al.}(2019)\citenamefont
  {Simonsen}, \citenamefont {S{\'{a}}nchez-Heredia}, \citenamefont {Saarinen},
  \citenamefont {Ardenkj{\ae}r-Larsen}, \citenamefont {Schliesser},\ and\
  \citenamefont {Polzik}}]{Simonsen2019}%
  \BibitemOpen
  \bibfield  {author} {\bibinfo {author} {\bibfnamefont {A.}~\bibnamefont
  {Simonsen}}, \bibinfo {author} {\bibfnamefont {J.~D.}\ \bibnamefont
  {S{\'{a}}nchez-Heredia}}, \bibinfo {author} {\bibfnamefont {S.~A.}\
  \bibnamefont {Saarinen}}, \bibinfo {author} {\bibfnamefont {J.~H.}\
  \bibnamefont {Ardenkj{\ae}r-Larsen}}, \bibinfo {author} {\bibfnamefont
  {A.}~\bibnamefont {Schliesser}},\ and\ \bibinfo {author} {\bibfnamefont
  {E.~S.}\ \bibnamefont {Polzik}},\ }\bibfield  {title} {\bibinfo {title}
  {Magnetic resonance imaging with optical preamplification and detection},\
  }\bibfield  {journal} {\bibinfo  {journal} {Scientific Reports}\ }\textbf
  {\bibinfo {volume} {9}},\ \href {https://doi.org/10.1038/s41598-019-54200-3}
  {10.1038/s41598-019-54200-3} (\bibinfo {year} {2019})\BibitemShut {NoStop}%
\bibitem [{\citenamefont {Piller}\ \emph {et~al.}(2019)\citenamefont {Piller},
  \citenamefont {Luhmann}, \citenamefont {Chien},\ and\ \citenamefont
  {Schmid}}]{piller2019nanoelectromechanical}%
  \BibitemOpen
  \bibfield  {author} {\bibinfo {author} {\bibfnamefont {M.}~\bibnamefont
  {Piller}}, \bibinfo {author} {\bibfnamefont {N.}~\bibnamefont {Luhmann}},
  \bibinfo {author} {\bibfnamefont {M.-H.}\ \bibnamefont {Chien}},\ and\
  \bibinfo {author} {\bibfnamefont {S.}~\bibnamefont {Schmid}},\ }\bibfield
  {title} {\bibinfo {title} {Nanoelectromechanical infrared detector},\ }in\
  \href {https://doi.org/https://doi.org/10.1117/12.2528416} {\emph {\bibinfo
  {booktitle} {Optical Sensing, Imaging, and Photon Counting: From X-Rays to
  THz 2019}}},\ Vol.\ \bibinfo {volume} {11088}\ (\bibinfo {organization}
  {International Society for Optics and Photonics},\ \bibinfo {year} {2019})\
  p.\ \bibinfo {pages} {1108802}\BibitemShut {NoStop}%
\bibitem [{\citenamefont {Yang}\ \emph {et~al.}(2019)\citenamefont {Yang},
  \citenamefont {Rochau}, \citenamefont {Huber}, \citenamefont {Brieussel},
  \citenamefont {Rastelli}, \citenamefont {Weig},\ and\ \citenamefont
  {Scheer}}]{Yang2019}%
  \BibitemOpen
  \bibfield  {author} {\bibinfo {author} {\bibfnamefont {F.}~\bibnamefont
  {Yang}}, \bibinfo {author} {\bibfnamefont {F.}~\bibnamefont {Rochau}},
  \bibinfo {author} {\bibfnamefont {J.~S.}\ \bibnamefont {Huber}}, \bibinfo
  {author} {\bibfnamefont {A.}~\bibnamefont {Brieussel}}, \bibinfo {author}
  {\bibfnamefont {G.}~\bibnamefont {Rastelli}}, \bibinfo {author}
  {\bibfnamefont {E.~M.}\ \bibnamefont {Weig}},\ and\ \bibinfo {author}
  {\bibfnamefont {E.}~\bibnamefont {Scheer}},\ }\bibfield  {title} {\bibinfo
  {title} {Spatial modulation of nonlinear flexural vibrations of membrane
  resonators},\ }\href@noop {} {\bibfield  {journal} {\bibinfo  {journal}
  {Physical review letters}\ }\textbf {\bibinfo {volume} {122}},\ \bibinfo
  {pages} {154301} (\bibinfo {year} {2019})}\BibitemShut {NoStop}%
\bibitem [{\citenamefont {Huber}\ \emph {et~al.}(2020)\citenamefont {Huber},
  \citenamefont {Rastelli}, \citenamefont {Seitner}, \citenamefont {K{\"o}lbl},
  \citenamefont {Belzig}, \citenamefont {Dykman},\ and\ \citenamefont
  {Weig}}]{huber2020spectral}%
  \BibitemOpen
  \bibfield  {author} {\bibinfo {author} {\bibfnamefont {J.~S.}\ \bibnamefont
  {Huber}}, \bibinfo {author} {\bibfnamefont {G.}~\bibnamefont {Rastelli}},
  \bibinfo {author} {\bibfnamefont {M.~J.}\ \bibnamefont {Seitner}}, \bibinfo
  {author} {\bibfnamefont {J.}~\bibnamefont {K{\"o}lbl}}, \bibinfo {author}
  {\bibfnamefont {W.}~\bibnamefont {Belzig}}, \bibinfo {author} {\bibfnamefont
  {M.~I.}\ \bibnamefont {Dykman}},\ and\ \bibinfo {author} {\bibfnamefont
  {E.~M.}\ \bibnamefont {Weig}},\ }\bibfield  {title} {\bibinfo {title}
  {Spectral evidence of squeezing of a weakly damped driven nanomechanical
  mode},\ }\href@noop {} {\bibfield  {journal} {\bibinfo  {journal} {Physical
  Review X}\ }\textbf {\bibinfo {volume} {10}},\ \bibinfo {pages} {021066}
  (\bibinfo {year} {2020})}\BibitemShut {NoStop}%
\bibitem [{\citenamefont {Karg}\ \emph {et~al.}(2020)\citenamefont {Karg},
  \citenamefont {Gouraud}, \citenamefont {Ngai}, \citenamefont {Schmid},
  \citenamefont {Hammerer},\ and\ \citenamefont {Treutlein}}]{Karg2020}%
  \BibitemOpen
  \bibfield  {author} {\bibinfo {author} {\bibfnamefont {T.~M.}\ \bibnamefont
  {Karg}}, \bibinfo {author} {\bibfnamefont {B.}~\bibnamefont {Gouraud}},
  \bibinfo {author} {\bibfnamefont {C.~T.}\ \bibnamefont {Ngai}}, \bibinfo
  {author} {\bibfnamefont {G.-L.}\ \bibnamefont {Schmid}}, \bibinfo {author}
  {\bibfnamefont {K.}~\bibnamefont {Hammerer}},\ and\ \bibinfo {author}
  {\bibfnamefont {P.}~\bibnamefont {Treutlein}},\ }\bibfield  {title} {\bibinfo
  {title} {Light-mediated strong coupling between a mechanical oscillator and
  atomic spins 1 meter apart},\ }\href
  {https://doi.org/10.1126/science.abb0328} {\bibfield  {journal} {\bibinfo
  {journal} {Science}\ }\textbf {\bibinfo {volume} {369}},\ \bibinfo {pages}
  {174} (\bibinfo {year} {2020})}\BibitemShut {NoStop}%
\bibitem [{\citenamefont {Sudhir}\ \emph {et~al.}(2017)\citenamefont {Sudhir},
  \citenamefont {Schilling}, \citenamefont {Fedorov}, \citenamefont
  {Sch{\"u}tz}, \citenamefont {Wilson},\ and\ \citenamefont
  {Kippenberg}}]{sudhir2017quantum}%
  \BibitemOpen
  \bibfield  {author} {\bibinfo {author} {\bibfnamefont {V.}~\bibnamefont
  {Sudhir}}, \bibinfo {author} {\bibfnamefont {R.}~\bibnamefont {Schilling}},
  \bibinfo {author} {\bibfnamefont {S.~A.}\ \bibnamefont {Fedorov}}, \bibinfo
  {author} {\bibfnamefont {H.}~\bibnamefont {Sch{\"u}tz}}, \bibinfo {author}
  {\bibfnamefont {D.~J.}\ \bibnamefont {Wilson}},\ and\ \bibinfo {author}
  {\bibfnamefont {T.~J.}\ \bibnamefont {Kippenberg}},\ }\bibfield  {title}
  {\bibinfo {title} {Quantum correlations of light from a room-temperature
  mechanical oscillator},\ }\href@noop {} {\bibfield  {journal} {\bibinfo
  {journal} {Physical Review X}\ }\textbf {\bibinfo {volume} {7}},\ \bibinfo
  {pages} {031055} (\bibinfo {year} {2017})}\BibitemShut {NoStop}%
\bibitem [{\citenamefont {Rossi}\ \emph {et~al.}(2018)\citenamefont {Rossi},
  \citenamefont {Mason}, \citenamefont {Chen}, \citenamefont {Tsaturyan},\ and\
  \citenamefont {Schliesser}}]{rossi2018measurement}%
  \BibitemOpen
  \bibfield  {author} {\bibinfo {author} {\bibfnamefont {M.}~\bibnamefont
  {Rossi}}, \bibinfo {author} {\bibfnamefont {D.}~\bibnamefont {Mason}},
  \bibinfo {author} {\bibfnamefont {J.}~\bibnamefont {Chen}}, \bibinfo {author}
  {\bibfnamefont {Y.}~\bibnamefont {Tsaturyan}},\ and\ \bibinfo {author}
  {\bibfnamefont {A.}~\bibnamefont {Schliesser}},\ }\bibfield  {title}
  {\bibinfo {title} {Measurement-based quantum control of mechanical motion},\
  }\href {https://doi.org/https://doi.org/10.1038/s41586-018-0643-8} {\bibfield
   {journal} {\bibinfo  {journal} {Nature}\ }\textbf {\bibinfo {volume}
  {563}},\ \bibinfo {pages} {53} (\bibinfo {year} {2018})}\BibitemShut
  {NoStop}%
\bibitem [{\citenamefont {Mason}\ \emph {et~al.}(2019)\citenamefont {Mason},
  \citenamefont {Chen}, \citenamefont {Rossi}, \citenamefont {Tsaturyan},\ and\
  \citenamefont {Schliesser}}]{mason2019continuous}%
  \BibitemOpen
  \bibfield  {author} {\bibinfo {author} {\bibfnamefont {D.}~\bibnamefont
  {Mason}}, \bibinfo {author} {\bibfnamefont {J.}~\bibnamefont {Chen}},
  \bibinfo {author} {\bibfnamefont {M.}~\bibnamefont {Rossi}}, \bibinfo
  {author} {\bibfnamefont {Y.}~\bibnamefont {Tsaturyan}},\ and\ \bibinfo
  {author} {\bibfnamefont {A.}~\bibnamefont {Schliesser}},\ }\bibfield  {title}
  {\bibinfo {title} {Continuous force and displacement measurement below the
  standard quantum limit},\ }\href
  {https://doi.org/https://doi.org/10.1038/nnano.2017.101} {\bibfield
  {journal} {\bibinfo  {journal} {Nature Physics}\ }\textbf {\bibinfo {volume}
  {15}},\ \bibinfo {pages} {745} (\bibinfo {year} {2019})}\BibitemShut
  {NoStop}%
\bibitem [{\citenamefont {Guo}\ \emph {et~al.}(2019)\citenamefont {Guo},
  \citenamefont {Norte},\ and\ \citenamefont {Gr{\"o}blacher}}]{Guo2019}%
  \BibitemOpen
  \bibfield  {author} {\bibinfo {author} {\bibfnamefont {J.}~\bibnamefont
  {Guo}}, \bibinfo {author} {\bibfnamefont {R.}~\bibnamefont {Norte}},\ and\
  \bibinfo {author} {\bibfnamefont {S.}~\bibnamefont {Gr{\"o}blacher}},\
  }\bibfield  {title} {\bibinfo {title} {Feedback cooling of a room temperature
  mechanical oscillator close to its motional ground state},\ }\href@noop {}
  {\bibfield  {journal} {\bibinfo  {journal} {Physical review letters}\
  }\textbf {\bibinfo {volume} {123}},\ \bibinfo {pages} {223602} (\bibinfo
  {year} {2019})}\BibitemShut {NoStop}%
\bibitem [{\citenamefont {Andrews}\ \emph {et~al.}(2014)\citenamefont
  {Andrews}, \citenamefont {Peterson}, \citenamefont {Purdy}, \citenamefont
  {Cicak}, \citenamefont {Simmonds}, \citenamefont {Regal},\ and\ \citenamefont
  {Lehnert}}]{Andrews2014}%
  \BibitemOpen
  \bibfield  {author} {\bibinfo {author} {\bibfnamefont {R.~W.}\ \bibnamefont
  {Andrews}}, \bibinfo {author} {\bibfnamefont {R.~W.}\ \bibnamefont
  {Peterson}}, \bibinfo {author} {\bibfnamefont {T.~P.}\ \bibnamefont {Purdy}},
  \bibinfo {author} {\bibfnamefont {K.}~\bibnamefont {Cicak}}, \bibinfo
  {author} {\bibfnamefont {R.~W.}\ \bibnamefont {Simmonds}}, \bibinfo {author}
  {\bibfnamefont {C.~A.}\ \bibnamefont {Regal}},\ and\ \bibinfo {author}
  {\bibfnamefont {K.~W.}\ \bibnamefont {Lehnert}},\ }\bibfield  {title}
  {\bibinfo {title} {Bidirectional and efficient conversion between microwave
  and optical light},\ }\href {https://doi.org/10.1038/nphys2911} {\bibfield
  {journal} {\bibinfo  {journal} {Nature Physics}\ }\textbf {\bibinfo {volume}
  {10}},\ \bibinfo {pages} {321} (\bibinfo {year} {2014})}\BibitemShut
  {NoStop}%
\bibitem [{\citenamefont {O'Connell}\ \emph {et~al.}(2010)\citenamefont
  {O'Connell}, \citenamefont {Hofheinz}, \citenamefont {Ansmann}, \citenamefont
  {Bialczak}, \citenamefont {Lenander}, \citenamefont {Lucero}, \citenamefont
  {Neeley}, \citenamefont {Sank}, \citenamefont {Wang}, \citenamefont {Weides},
  \citenamefont {Wenner}, \citenamefont {Martinis},\ and\ \citenamefont
  {Cleland}}]{OConnell2010}%
  \BibitemOpen
  \bibfield  {author} {\bibinfo {author} {\bibfnamefont {A.~D.}\ \bibnamefont
  {O'Connell}}, \bibinfo {author} {\bibfnamefont {M.}~\bibnamefont {Hofheinz}},
  \bibinfo {author} {\bibfnamefont {M.}~\bibnamefont {Ansmann}}, \bibinfo
  {author} {\bibfnamefont {R.~C.}\ \bibnamefont {Bialczak}}, \bibinfo {author}
  {\bibfnamefont {M.}~\bibnamefont {Lenander}}, \bibinfo {author}
  {\bibfnamefont {E.}~\bibnamefont {Lucero}}, \bibinfo {author} {\bibfnamefont
  {M.}~\bibnamefont {Neeley}}, \bibinfo {author} {\bibfnamefont
  {D.}~\bibnamefont {Sank}}, \bibinfo {author} {\bibfnamefont {H.}~\bibnamefont
  {Wang}}, \bibinfo {author} {\bibfnamefont {M.}~\bibnamefont {Weides}},
  \bibinfo {author} {\bibfnamefont {J.}~\bibnamefont {Wenner}}, \bibinfo
  {author} {\bibfnamefont {J.~M.}\ \bibnamefont {Martinis}},\ and\ \bibinfo
  {author} {\bibfnamefont {A.~N.}\ \bibnamefont {Cleland}},\ }\bibfield
  {title} {\bibinfo {title} {Quantum ground state and single-phonon control of
  a mechanical resonator},\ }\href {https://doi.org/10.1038/nature08967}
  {\bibfield  {journal} {\bibinfo  {journal} {Nature}\ }\textbf {\bibinfo
  {volume} {464}},\ \bibinfo {pages} {697} (\bibinfo {year}
  {2010})}\BibitemShut {NoStop}%
\bibitem [{\citenamefont {Thompson}\ \emph {et~al.}(2008)\citenamefont
  {Thompson}, \citenamefont {Zwickl}, \citenamefont {Jayich}, \citenamefont
  {Marquardt}, \citenamefont {Girvin},\ and\ \citenamefont
  {Harris}}]{Thompson2008}%
  \BibitemOpen
  \bibfield  {author} {\bibinfo {author} {\bibfnamefont {J.~D.}\ \bibnamefont
  {Thompson}}, \bibinfo {author} {\bibfnamefont {B.~M.}\ \bibnamefont
  {Zwickl}}, \bibinfo {author} {\bibfnamefont {A.~M.}\ \bibnamefont {Jayich}},
  \bibinfo {author} {\bibfnamefont {F.}~\bibnamefont {Marquardt}}, \bibinfo
  {author} {\bibfnamefont {S.~M.}\ \bibnamefont {Girvin}},\ and\ \bibinfo
  {author} {\bibfnamefont {J.~G.~E.}\ \bibnamefont {Harris}},\ }\bibfield
  {title} {\bibinfo {title} {Strong dispersive coupling of a high-finesse
  cavity to a micromechanical membrane},\ }\href
  {https://doi.org/10.1038/nature06715} {\bibfield  {journal} {\bibinfo
  {journal} {Nature}\ }\textbf {\bibinfo {volume} {452}},\ \bibinfo {pages}
  {72} (\bibinfo {year} {2008})}\BibitemShut {NoStop}%
\bibitem [{\citenamefont {Wilson}\ \emph {et~al.}(2009)\citenamefont {Wilson},
  \citenamefont {Regal}, \citenamefont {Papp},\ and\ \citenamefont
  {Kimble}}]{Wilson2009}%
  \BibitemOpen
  \bibfield  {author} {\bibinfo {author} {\bibfnamefont {D.~J.}\ \bibnamefont
  {Wilson}}, \bibinfo {author} {\bibfnamefont {C.~A.}\ \bibnamefont {Regal}},
  \bibinfo {author} {\bibfnamefont {S.~B.}\ \bibnamefont {Papp}},\ and\
  \bibinfo {author} {\bibfnamefont {H.~J.}\ \bibnamefont {Kimble}},\ }\bibfield
   {title} {\bibinfo {title} {Cavity optomechanics with stoichiometric sin
  films},\ }\href {https://doi.org/10.1103/PhysRevLett.103.207204} {\bibfield
  {journal} {\bibinfo  {journal} {Physical Review Letters}\ }\textbf {\bibinfo
  {volume} {103}},\ \bibinfo {eid} {207204} (\bibinfo {year}
  {2009})}\BibitemShut {NoStop}%
\bibitem [{\citenamefont {J{\"o}ckel}\ \emph {et~al.}(2014)\citenamefont
  {J{\"o}ckel}, \citenamefont {Faber}, \citenamefont {Kampschulte},
  \citenamefont {Korppi}, \citenamefont {Rakher},\ and\ \citenamefont
  {Treutlein}}]{Joeckel2014}%
  \BibitemOpen
  \bibfield  {author} {\bibinfo {author} {\bibfnamefont {A.}~\bibnamefont
  {J{\"o}ckel}}, \bibinfo {author} {\bibfnamefont {A.}~\bibnamefont {Faber}},
  \bibinfo {author} {\bibfnamefont {T.}~\bibnamefont {Kampschulte}}, \bibinfo
  {author} {\bibfnamefont {M.}~\bibnamefont {Korppi}}, \bibinfo {author}
  {\bibfnamefont {M.~T.}\ \bibnamefont {Rakher}},\ and\ \bibinfo {author}
  {\bibfnamefont {P.}~\bibnamefont {Treutlein}},\ }\bibfield  {title} {\bibinfo
  {title} {Sympathetic cooling of a membrane oscillator in a hybrid mechanical
  atomic system},\ }\href {https://doi.org/10.1038/nnano.2014.278} {\bibfield
  {journal} {\bibinfo  {journal} {Nature Nanotechnology}\ }\textbf {\bibinfo
  {volume} {10}},\ \bibinfo {pages} {55} (\bibinfo {year} {2014})}\BibitemShut
  {NoStop}%
\bibitem [{\citenamefont {Ftouni}\ \emph {et~al.}(2015)\citenamefont {Ftouni},
  \citenamefont {Blanc}, \citenamefont {Tainoff}, \citenamefont {Fefferman},
  \citenamefont {Defoort}, \citenamefont {Lulla}, \citenamefont {Richard},
  \citenamefont {Collin},\ and\ \citenamefont {Bourgeois}}]{Ftouni2015}%
  \BibitemOpen
  \bibfield  {author} {\bibinfo {author} {\bibfnamefont {H.}~\bibnamefont
  {Ftouni}}, \bibinfo {author} {\bibfnamefont {C.}~\bibnamefont {Blanc}},
  \bibinfo {author} {\bibfnamefont {D.}~\bibnamefont {Tainoff}}, \bibinfo
  {author} {\bibfnamefont {A.~D.}\ \bibnamefont {Fefferman}}, \bibinfo {author}
  {\bibfnamefont {M.}~\bibnamefont {Defoort}}, \bibinfo {author} {\bibfnamefont
  {K.~J.}\ \bibnamefont {Lulla}}, \bibinfo {author} {\bibfnamefont
  {J.}~\bibnamefont {Richard}}, \bibinfo {author} {\bibfnamefont
  {E.}~\bibnamefont {Collin}},\ and\ \bibinfo {author} {\bibfnamefont
  {O.}~\bibnamefont {Bourgeois}},\ }\bibfield  {title} {\bibinfo {title}
  {Thermal conductivity of silicon nitride membranes is not sensitive to
  stress},\ }\href {https://doi.org/10.1103/PhysRevB.92.125439} {\bibfield
  {journal} {\bibinfo  {journal} {Phys. Rev. B}\ }\textbf {\bibinfo {volume}
  {92}},\ \bibinfo {pages} {125439} (\bibinfo {year} {2015})}\BibitemShut
  {NoStop}%
\bibitem [{\citenamefont {Fink}\ \emph {et~al.}(2016)\citenamefont {Fink},
  \citenamefont {Kalaee}, \citenamefont {Pitanti}, \citenamefont {Norte},
  \citenamefont {Heinzle}, \citenamefont {Davan{\c{c}}o}, \citenamefont
  {Srinivasan},\ and\ \citenamefont {Painter}}]{Fink2016}%
  \BibitemOpen
  \bibfield  {author} {\bibinfo {author} {\bibfnamefont {J.~M.}\ \bibnamefont
  {Fink}}, \bibinfo {author} {\bibfnamefont {M.}~\bibnamefont {Kalaee}},
  \bibinfo {author} {\bibfnamefont {A.}~\bibnamefont {Pitanti}}, \bibinfo
  {author} {\bibfnamefont {R.}~\bibnamefont {Norte}}, \bibinfo {author}
  {\bibfnamefont {L.}~\bibnamefont {Heinzle}}, \bibinfo {author} {\bibfnamefont
  {M.}~\bibnamefont {Davan{\c{c}}o}}, \bibinfo {author} {\bibfnamefont
  {K.}~\bibnamefont {Srinivasan}},\ and\ \bibinfo {author} {\bibfnamefont
  {O.}~\bibnamefont {Painter}},\ }\bibfield  {title} {\bibinfo {title} {Quantum
  electromechanics on silicon nitride nanomembranes},\ }\bibfield  {journal}
  {\bibinfo  {journal} {Nature Communications}\ }\textbf {\bibinfo {volume}
  {7}},\ \href {https://doi.org/10.1038/ncomms12396} {10.1038/ncomms12396}
  (\bibinfo {year} {2016})\BibitemShut {NoStop}%
\bibitem [{\citenamefont {Serra}\ \emph {et~al.}(2016)\citenamefont {Serra},
  \citenamefont {Bawaj}, \citenamefont {Borrielli}, \citenamefont {Giuseppe},
  \citenamefont {Forte}, \citenamefont {Kralj}, \citenamefont {Malossi},
  \citenamefont {Marconi}, \citenamefont {Marin}, \citenamefont {Marino},
  \citenamefont {Morana}, \citenamefont {Natali}, \citenamefont {Pandraud},
  \citenamefont {Pontin}, \citenamefont {Prodi}, \citenamefont {Rossi},
  \citenamefont {Sarro}, \citenamefont {Vitali},\ and\ \citenamefont
  {Bonaldi}}]{Serra2016}%
  \BibitemOpen
  \bibfield  {author} {\bibinfo {author} {\bibfnamefont {E.}~\bibnamefont
  {Serra}}, \bibinfo {author} {\bibfnamefont {M.}~\bibnamefont {Bawaj}},
  \bibinfo {author} {\bibfnamefont {A.}~\bibnamefont {Borrielli}}, \bibinfo
  {author} {\bibfnamefont {G.~D.}\ \bibnamefont {Giuseppe}}, \bibinfo {author}
  {\bibfnamefont {S.}~\bibnamefont {Forte}}, \bibinfo {author} {\bibfnamefont
  {N.}~\bibnamefont {Kralj}}, \bibinfo {author} {\bibfnamefont
  {N.}~\bibnamefont {Malossi}}, \bibinfo {author} {\bibfnamefont
  {L.}~\bibnamefont {Marconi}}, \bibinfo {author} {\bibfnamefont
  {F.}~\bibnamefont {Marin}}, \bibinfo {author} {\bibfnamefont
  {F.}~\bibnamefont {Marino}}, \bibinfo {author} {\bibfnamefont
  {B.}~\bibnamefont {Morana}}, \bibinfo {author} {\bibfnamefont
  {R.}~\bibnamefont {Natali}}, \bibinfo {author} {\bibfnamefont
  {G.}~\bibnamefont {Pandraud}}, \bibinfo {author} {\bibfnamefont
  {A.}~\bibnamefont {Pontin}}, \bibinfo {author} {\bibfnamefont {G.~A.}\
  \bibnamefont {Prodi}}, \bibinfo {author} {\bibfnamefont {M.}~\bibnamefont
  {Rossi}}, \bibinfo {author} {\bibfnamefont {P.~M.}\ \bibnamefont {Sarro}},
  \bibinfo {author} {\bibfnamefont {D.}~\bibnamefont {Vitali}},\ and\ \bibinfo
  {author} {\bibfnamefont {M.}~\bibnamefont {Bonaldi}},\ }\bibfield  {title}
  {\bibinfo {title} {Microfabrication of large-area circular high-stress
  silicon nitride membranes for optomechanical applications},\ }\href
  {https://doi.org/10.1063/1.4953805} {\bibfield  {journal} {\bibinfo
  {journal} {{AIP} Advances}\ }\textbf {\bibinfo {volume} {6}},\ \bibinfo
  {pages} {065004} (\bibinfo {year} {2016})}\BibitemShut {NoStop}%
\bibitem [{\citenamefont {Norte}\ \emph {et~al.}(2016)\citenamefont {Norte},
  \citenamefont {Moura},\ and\ \citenamefont {Gr{\"o}blacher}}]{Norte2016}%
  \BibitemOpen
  \bibfield  {author} {\bibinfo {author} {\bibfnamefont {R.~A.}\ \bibnamefont
  {Norte}}, \bibinfo {author} {\bibfnamefont {J.~P.}\ \bibnamefont {Moura}},\
  and\ \bibinfo {author} {\bibfnamefont {S.}~\bibnamefont {Gr{\"o}blacher}},\
  }\bibfield  {title} {\bibinfo {title} {Mechanical resonators for quantum
  optomechanics experiments at room temperature},\ }\href
  {https://doi.org/10.1103/PhysRevLett.116.147202} {\bibfield  {journal}
  {\bibinfo  {journal} {Phys. Rev. Lett.}\ }\textbf {\bibinfo {volume} {116}},\
  \bibinfo {pages} {147202} (\bibinfo {year} {2016})}\BibitemShut {NoStop}%
\bibitem [{\citenamefont {Reinhardt}\ \emph {et~al.}(2016)\citenamefont
  {Reinhardt}, \citenamefont {M\"uller}, \citenamefont {Bourassa},\ and\
  \citenamefont {Sankey}}]{Reinhardt2016}%
  \BibitemOpen
  \bibfield  {author} {\bibinfo {author} {\bibfnamefont {C.}~\bibnamefont
  {Reinhardt}}, \bibinfo {author} {\bibfnamefont {T.}~\bibnamefont {M\"uller}},
  \bibinfo {author} {\bibfnamefont {A.}~\bibnamefont {Bourassa}},\ and\
  \bibinfo {author} {\bibfnamefont {J.~C.}\ \bibnamefont {Sankey}},\ }\bibfield
   {title} {\bibinfo {title} {Ultralow-noise sin trampoline resonators for
  sensing and optomechanics},\ }\href
  {https://doi.org/10.1103/PhysRevX.6.021001} {\bibfield  {journal} {\bibinfo
  {journal} {Phys. Rev. X}\ }\textbf {\bibinfo {volume} {6}},\ \bibinfo {pages}
  {021001} (\bibinfo {year} {2016})}\BibitemShut {NoStop}%
\bibitem [{\citenamefont {{Tsaturyan}}\ \emph {et~al.}(2017)\citenamefont
  {{Tsaturyan}}, \citenamefont {{Barg}}, \citenamefont {{Polzik}},\ and\
  \citenamefont
  {{Schliesser}}}]{Tsaturyan2017NatNano-UltracoherentNanomechSoftclamping}%
  \BibitemOpen
  \bibfield  {author} {\bibinfo {author} {\bibfnamefont {Y.}~\bibnamefont
  {{Tsaturyan}}}, \bibinfo {author} {\bibfnamefont {A.}~\bibnamefont {{Barg}}},
  \bibinfo {author} {\bibfnamefont {E.~S.}\ \bibnamefont {{Polzik}}},\ and\
  \bibinfo {author} {\bibfnamefont {A.}~\bibnamefont {{Schliesser}}},\
  }\bibfield  {title} {\bibinfo {title} {{Ultracoherent nanomechanical
  resonators via soft clamping and dissipation dilution}},\ }\href
  {https://doi.org/10.1038/nnano.2017.101} {\bibfield  {journal} {\bibinfo
  {journal} {Nature Nanotechnology}\ }\textbf {\bibinfo {volume} {12}},\
  \bibinfo {pages} {776} (\bibinfo {year} {2017})}\BibitemShut {NoStop}%
\bibitem [{\citenamefont {{Ghadimi}}\ \emph {et~al.}(2018)\citenamefont
  {{Ghadimi}}, \citenamefont {{Fedorov}}, \citenamefont {{Engelsen}},
  \citenamefont {{Bereyhi}}, \citenamefont {{Schilling}}, \citenamefont
  {{Wilson}},\ and\ \citenamefont
  {{Kippenberg}}}]{Ghadimi2018StrainEngineering}%
  \BibitemOpen
  \bibfield  {author} {\bibinfo {author} {\bibfnamefont {A.~H.}\ \bibnamefont
  {{Ghadimi}}}, \bibinfo {author} {\bibfnamefont {S.~A.}\ \bibnamefont
  {{Fedorov}}}, \bibinfo {author} {\bibfnamefont {N.~J.}\ \bibnamefont
  {{Engelsen}}}, \bibinfo {author} {\bibfnamefont {M.~J.}\ \bibnamefont
  {{Bereyhi}}}, \bibinfo {author} {\bibfnamefont {R.}~\bibnamefont
  {{Schilling}}}, \bibinfo {author} {\bibfnamefont {D.~J.}\ \bibnamefont
  {{Wilson}}},\ and\ \bibinfo {author} {\bibfnamefont {T.~J.}\ \bibnamefont
  {{Kippenberg}}},\ }\bibfield  {title} {\bibinfo {title} {{Elastic strain
  engineering for ultralow mechanical dissipation}},\ }\href
  {https://doi.org/10.1126/science.aar6939} {\bibfield  {journal} {\bibinfo
  {journal} {Science}\ }\textbf {\bibinfo {volume} {360}},\ \bibinfo {pages}
  {764} (\bibinfo {year} {2018})}\BibitemShut {NoStop}%
\bibitem [{\citenamefont {Yuksel}\ \emph {et~al.}(2019)\citenamefont {Yuksel},
  \citenamefont {Orhan}, \citenamefont {Yanik}, \citenamefont {Ari},
  \citenamefont {Demir},\ and\ \citenamefont {Hanay}}]{yuksel2019nonlinear}%
  \BibitemOpen
  \bibfield  {author} {\bibinfo {author} {\bibfnamefont {M.}~\bibnamefont
  {Yuksel}}, \bibinfo {author} {\bibfnamefont {E.}~\bibnamefont {Orhan}},
  \bibinfo {author} {\bibfnamefont {C.}~\bibnamefont {Yanik}}, \bibinfo
  {author} {\bibfnamefont {A.~B.}\ \bibnamefont {Ari}}, \bibinfo {author}
  {\bibfnamefont {A.}~\bibnamefont {Demir}},\ and\ \bibinfo {author}
  {\bibfnamefont {M.~S.}\ \bibnamefont {Hanay}},\ }\bibfield  {title} {\bibinfo
  {title} {Nonlinear nanomechanical mass spectrometry at the
  single-nanoparticle level},\ }\href@noop {} {\bibfield  {journal} {\bibinfo
  {journal} {Nano letters}\ }\textbf {\bibinfo {volume} {19}},\ \bibinfo
  {pages} {3583} (\bibinfo {year} {2019})}\BibitemShut {NoStop}%
\bibitem [{\citenamefont {{Gonz{\'a}lez}}\ and\ \citenamefont
  {{Saulson}}(1994)}]{GonzalezSaulson1994BrownianMotionAnelasticWire}%
  \BibitemOpen
  \bibfield  {author} {\bibinfo {author} {\bibfnamefont {G.~I.}\ \bibnamefont
  {{Gonz{\'a}lez}}}\ and\ \bibinfo {author} {\bibfnamefont {P.~R.}\
  \bibnamefont {{Saulson}}},\ }\bibfield  {title} {\bibinfo {title} {{Brownian
  motion of a mass suspended by an anelastic wire}},\ }\href
  {https://doi.org/10.1121/1.410467} {\bibfield  {journal} {\bibinfo  {journal}
  {Acoustical Society of America Journal}\ }\textbf {\bibinfo {volume} {96}},\
  \bibinfo {pages} {207} (\bibinfo {year} {1994})}\BibitemShut {NoStop}%
\bibitem [{\citenamefont {{Unterreithmeier}}\ \emph {et~al.}(2010)\citenamefont
  {{Unterreithmeier}}, \citenamefont {{Faust}},\ and\ \citenamefont
  {{Kotthaus}}}]{Unterreithmeier2010PRL-DampingNanomechRes}%
  \BibitemOpen
  \bibfield  {author} {\bibinfo {author} {\bibfnamefont {Q.~P.}\ \bibnamefont
  {{Unterreithmeier}}}, \bibinfo {author} {\bibfnamefont {T.}~\bibnamefont
  {{Faust}}},\ and\ \bibinfo {author} {\bibfnamefont {J.~P.}\ \bibnamefont
  {{Kotthaus}}},\ }\bibfield  {title} {\bibinfo {title} {{Damping of
  Nanomechanical Resonators}},\ }\href
  {https://doi.org/10.1103/PhysRevLett.105.027205} {\bibfield  {journal}
  {\bibinfo  {journal} {Phys. Rev. Lett.}\ }\textbf {\bibinfo {volume} {105}},\
  \bibinfo {eid} {027205} (\bibinfo {year} {2010})},\ \Eprint
  {https://arxiv.org/abs/1003.1868} {arXiv:1003.1868 [cond-mat.mes-hall]}
  \BibitemShut {NoStop}%
\bibitem [{\citenamefont {{Yu}}\ \emph {et~al.}(2012)\citenamefont {{Yu}},
  \citenamefont {{Purdy}},\ and\ \citenamefont
  {{Regal}}}]{Yu2012PRL-ControlMaterialDampingHighQMembraneMicrores}%
  \BibitemOpen
  \bibfield  {author} {\bibinfo {author} {\bibfnamefont {P.-L.}\ \bibnamefont
  {{Yu}}}, \bibinfo {author} {\bibfnamefont {T.~P.}\ \bibnamefont {{Purdy}}},\
  and\ \bibinfo {author} {\bibfnamefont {C.~A.}\ \bibnamefont {{Regal}}},\
  }\bibfield  {title} {\bibinfo {title} {{Control of Material Damping in High-Q
  Membrane Microresonators}},\ }\href
  {https://doi.org/10.1103/PhysRevLett.108.083603} {\bibfield  {journal}
  {\bibinfo  {journal} {Phys. Rev. Lett.}\ }\textbf {\bibinfo {volume} {108}},\
  \bibinfo {eid} {083603} (\bibinfo {year} {2012})}\BibitemShut {NoStop}%
\bibitem [{\citenamefont {{Villanueva}}\ and\ \citenamefont
  {{Schmid}}(2014)}]{Villanueva2014PRL-EvidenceSurfaceLossSiN}%
  \BibitemOpen
  \bibfield  {author} {\bibinfo {author} {\bibfnamefont {L.~G.}\ \bibnamefont
  {{Villanueva}}}\ and\ \bibinfo {author} {\bibfnamefont {S.}~\bibnamefont
  {{Schmid}}},\ }\bibfield  {title} {\bibinfo {title} {{Evidence of Surface
  Loss as Ubiquitous Limiting Damping Mechanism in SiN Micro- and
  Nanomechanical Resonators}},\ }\href
  {https://doi.org/10.1103/PhysRevLett.113.227201} {\bibfield  {journal}
  {\bibinfo  {journal} {Phys. Rev. Lett.}\ }\textbf {\bibinfo {volume} {113}},\
  \bibinfo {eid} {227201} (\bibinfo {year} {2014})}\BibitemShut {NoStop}%
\bibitem [{\citenamefont {{Rieger}}\ \emph {et~al.}(2014)\citenamefont
  {{Rieger}}, \citenamefont {{Isacsson}}, \citenamefont {{Seitner}},
  \citenamefont {{Kotthaus}},\ and\ \citenamefont {{Weig}}}]{bib:Rieger2014}%
  \BibitemOpen
  \bibfield  {author} {\bibinfo {author} {\bibfnamefont {J.}~\bibnamefont
  {{Rieger}}}, \bibinfo {author} {\bibfnamefont {A.}~\bibnamefont
  {{Isacsson}}}, \bibinfo {author} {\bibfnamefont {M.~J.}\ \bibnamefont
  {{Seitner}}}, \bibinfo {author} {\bibfnamefont {J.~P.}\ \bibnamefont
  {{Kotthaus}}},\ and\ \bibinfo {author} {\bibfnamefont {E.~M.}\ \bibnamefont
  {{Weig}}},\ }\bibfield  {title} {\bibinfo {title} {{Energy losses of
  nanomechanical resonators induced by atomic force microscopy-controlled
  mechanical impedance mismatching}},\ }\href
  {https://doi.org/10.1038/ncomms4345} {\bibfield  {journal} {\bibinfo
  {journal} {Nature Communications}\ }\textbf {\bibinfo {volume} {5}},\
  \bibinfo {eid} {3345} (\bibinfo {year} {2014})}\BibitemShut {NoStop}%
\bibitem [{\citenamefont {Kleckner}\ \emph {et~al.}(2011)\citenamefont
  {Kleckner}, \citenamefont {Pepper}, \citenamefont {Jeffrey}, \citenamefont
  {Sonin}, \citenamefont {Thon},\ and\ \citenamefont
  {Bouwmeester}}]{Kleckner2011}%
  \BibitemOpen
  \bibfield  {author} {\bibinfo {author} {\bibfnamefont {D.}~\bibnamefont
  {Kleckner}}, \bibinfo {author} {\bibfnamefont {B.}~\bibnamefont {Pepper}},
  \bibinfo {author} {\bibfnamefont {E.}~\bibnamefont {Jeffrey}}, \bibinfo
  {author} {\bibfnamefont {P.}~\bibnamefont {Sonin}}, \bibinfo {author}
  {\bibfnamefont {S.~M.}\ \bibnamefont {Thon}},\ and\ \bibinfo {author}
  {\bibfnamefont {D.}~\bibnamefont {Bouwmeester}},\ }\bibfield  {title}
  {\bibinfo {title} {Optomechanical trampoline resonators},\ }\href
  {https://doi.org/10.1364/oe.19.019708} {\bibfield  {journal} {\bibinfo
  {journal} {Optics Express}\ }\textbf {\bibinfo {volume} {19}},\ \bibinfo
  {pages} {19708} (\bibinfo {year} {2011})}\BibitemShut {NoStop}%
\bibitem [{\citenamefont {Alegre}\ \emph {et~al.}(2011)\citenamefont {Alegre},
  \citenamefont {Safavi-Naeini}, \citenamefont {Winger},\ and\ \citenamefont
  {Painter}}]{Alegre2011}%
  \BibitemOpen
  \bibfield  {author} {\bibinfo {author} {\bibfnamefont {T.~P.~M.}\
  \bibnamefont {Alegre}}, \bibinfo {author} {\bibfnamefont {A.}~\bibnamefont
  {Safavi-Naeini}}, \bibinfo {author} {\bibfnamefont {M.}~\bibnamefont
  {Winger}},\ and\ \bibinfo {author} {\bibfnamefont {O.}~\bibnamefont
  {Painter}},\ }\bibfield  {title} {\bibinfo {title} {Quasi-two-dimensional
  optomechanical crystals with a complete phononicbandgap},\ }\href
  {https://doi.org/10.1364/OE.19.005658} {\bibfield  {journal} {\bibinfo
  {journal} {Opt. Express}\ }\textbf {\bibinfo {volume} {19}},\ \bibinfo
  {pages} {5658} (\bibinfo {year} {2011})}\BibitemShut {NoStop}%
\bibitem [{\citenamefont {Yu}\ \emph {et~al.}(2014)\citenamefont {Yu},
  \citenamefont {Cicak}, \citenamefont {Kampel}, \citenamefont {Tsaturyan},
  \citenamefont {Purdy}, \citenamefont {Simmonds},\ and\ \citenamefont
  {Regal}}]{yu2014phononic}%
  \BibitemOpen
  \bibfield  {author} {\bibinfo {author} {\bibfnamefont {P.-L.}\ \bibnamefont
  {Yu}}, \bibinfo {author} {\bibfnamefont {K.}~\bibnamefont {Cicak}}, \bibinfo
  {author} {\bibfnamefont {N.}~\bibnamefont {Kampel}}, \bibinfo {author}
  {\bibfnamefont {Y.}~\bibnamefont {Tsaturyan}}, \bibinfo {author}
  {\bibfnamefont {T.}~\bibnamefont {Purdy}}, \bibinfo {author} {\bibfnamefont
  {R.}~\bibnamefont {Simmonds}},\ and\ \bibinfo {author} {\bibfnamefont
  {C.}~\bibnamefont {Regal}},\ }\bibfield  {title} {\bibinfo {title} {A
  phononic bandgap shield for high-q membrane microresonators},\ }\href
  {https://doi.org/https://doi.org/10.1063/1.4862031} {\bibfield  {journal}
  {\bibinfo  {journal} {Applied Physics Letters}\ }\textbf {\bibinfo {volume}
  {104}},\ \bibinfo {pages} {023510} (\bibinfo {year} {2014})}\BibitemShut
  {NoStop}%
\bibitem [{\citenamefont {Tsaturyan}\ \emph {et~al.}(2014)\citenamefont
  {Tsaturyan}, \citenamefont {Barg}, \citenamefont {Simonsen}, \citenamefont
  {Villanueva}, \citenamefont {Schmid}, \citenamefont {Schliesser},\ and\
  \citenamefont {Polzik}}]{Tsaturyan2014}%
  \BibitemOpen
  \bibfield  {author} {\bibinfo {author} {\bibfnamefont {Y.}~\bibnamefont
  {Tsaturyan}}, \bibinfo {author} {\bibfnamefont {A.}~\bibnamefont {Barg}},
  \bibinfo {author} {\bibfnamefont {A.}~\bibnamefont {Simonsen}}, \bibinfo
  {author} {\bibfnamefont {L.~G.}\ \bibnamefont {Villanueva}}, \bibinfo
  {author} {\bibfnamefont {S.}~\bibnamefont {Schmid}}, \bibinfo {author}
  {\bibfnamefont {A.}~\bibnamefont {Schliesser}},\ and\ \bibinfo {author}
  {\bibfnamefont {E.~S.}\ \bibnamefont {Polzik}},\ }\bibfield  {title}
  {\bibinfo {title} {Demonstration of suppressed phonon tunneling losses in
  phononic bandgap shielded membrane resonators for high-q optomechanics},\
  }\href {https://doi.org/10.1364/OE.22.006810} {\bibfield  {journal} {\bibinfo
   {journal} {Opt. Express}\ }\textbf {\bibinfo {volume} {22}},\ \bibinfo
  {pages} {6810} (\bibinfo {year} {2014})}\BibitemShut {NoStop}%
\bibitem [{\citenamefont {Fedorov}\ \emph {et~al.}(2019)\citenamefont
  {Fedorov}, \citenamefont {Engelsen}, \citenamefont {Ghadimi}, \citenamefont
  {Bereyhi}, \citenamefont {Schilling}, \citenamefont {Wilson},\ and\
  \citenamefont {Kippenberg}}]{fedorov2019generalized}%
  \BibitemOpen
  \bibfield  {author} {\bibinfo {author} {\bibfnamefont {S.~A.}\ \bibnamefont
  {Fedorov}}, \bibinfo {author} {\bibfnamefont {N.~J.}\ \bibnamefont
  {Engelsen}}, \bibinfo {author} {\bibfnamefont {A.~H.}\ \bibnamefont
  {Ghadimi}}, \bibinfo {author} {\bibfnamefont {M.~J.}\ \bibnamefont
  {Bereyhi}}, \bibinfo {author} {\bibfnamefont {R.}~\bibnamefont {Schilling}},
  \bibinfo {author} {\bibfnamefont {D.~J.}\ \bibnamefont {Wilson}},\ and\
  \bibinfo {author} {\bibfnamefont {T.~J.}\ \bibnamefont {Kippenberg}},\
  }\bibfield  {title} {\bibinfo {title} {Generalized dissipation dilution in
  strained mechanical resonators},\ }\href
  {https://doi.org/https://doi.org/10.1103/PhysRevB.99.054107} {\bibfield
  {journal} {\bibinfo  {journal} {Physical Review B}\ }\textbf {\bibinfo
  {volume} {99}},\ \bibinfo {pages} {054107} (\bibinfo {year}
  {2019})}\BibitemShut {NoStop}%
\bibitem [{siC()}]{siCite}%
  \BibitemOpen
  \href@noop {} {\bibinfo {title} {See supplemental material at [url will be
  inserted by publisher] for additional wafer spezifications, material
  parameters, finite element method simulations, and a measurement showing the
  pedestal contraction}}\BibitemShut {NoStop}%
\bibitem [{\citenamefont {{Weaver Jr.}}\ \emph {et~al.}(1990)\citenamefont
  {{Weaver Jr.}}, \citenamefont {{Timoshenko}},\ and\ \citenamefont
  {{Young}}}]{Timoshenko1990-VibrationProblemsEngineering}%
  \BibitemOpen
  \bibfield  {author} {\bibinfo {author} {\bibfnamefont {W.}~\bibnamefont
  {{Weaver Jr.}}}, \bibinfo {author} {\bibfnamefont {S.~P.}\ \bibnamefont
  {{Timoshenko}}},\ and\ \bibinfo {author} {\bibfnamefont {D.~H.}\ \bibnamefont
  {{Young}}},\ }\href@noop {} {\emph {\bibinfo {title} {Vibration problems in
  engineering}}}\ (\bibinfo  {publisher} {Wiley},\ \bibinfo {address} {New
  York},\ \bibinfo {year} {1990})\BibitemShut {NoStop}%
\bibitem [{\citenamefont {Cleland}(2002)}]{Cleland2002foundations}%
  \BibitemOpen
  \bibfield  {author} {\bibinfo {author} {\bibfnamefont {A.~N.}\ \bibnamefont
  {Cleland}},\ }\href@noop {} {\emph {\bibinfo {title} {Foundations of
  Nanomechanics: From Solid-State Theory to Device Applications}}}\ (\bibinfo
  {publisher} {Springer},\ \bibinfo {address} {Berlin},\ \bibinfo {year}
  {2002})\BibitemShut {NoStop}%
\bibitem [{\citenamefont {Doster}()}]{DosterUnpub}%
  \BibitemOpen
  \bibfield  {author} {\bibinfo {author} {\bibfnamefont {J.}~\bibnamefont
  {Doster}},\ }\href@noop {} {\bibinfo  {journal} {unpublished}\ }\BibitemShut
  {NoStop}%
\bibitem [{\citenamefont {Maluf}\ and\ \citenamefont
  {Williams}(2004)}]{maluf2002introduction}%
  \BibitemOpen
\bibfield  {journal} {  }\bibfield  {author} {\bibinfo {author} {\bibfnamefont
  {N.}~\bibnamefont {Maluf}}\ and\ \bibinfo {author} {\bibfnamefont
  {K.}~\bibnamefont {Williams}},\ }\href@noop {} {\emph {\bibinfo {title} {An
  introduction to microelectromechanical systems engineering}}}\ (\bibinfo
  {publisher} {Artech House},\ \bibinfo {address} {Norwood},\ \bibinfo {year}
  {2004})\BibitemShut {NoStop}%
\bibitem [{\citenamefont {Li}\ and\ \citenamefont
  {Bradt}(1987)}]{li1987single}%
  \BibitemOpen
  \bibfield  {author} {\bibinfo {author} {\bibfnamefont {Z.}~\bibnamefont
  {Li}}\ and\ \bibinfo {author} {\bibfnamefont {R.~C.}\ \bibnamefont {Bradt}},\
  }\bibfield  {title} {\bibinfo {title} {The single-crystal elastic constants
  of cubic (3c) sic to 1000 c},\ }\href@noop {} {\bibfield  {journal} {\bibinfo
   {journal} {Journal of materials science}\ }\textbf {\bibinfo {volume}
  {22}},\ \bibinfo {pages} {2557} (\bibinfo {year} {1987})}\BibitemShut
  {NoStop}%
\bibitem [{\citenamefont {Henisch}\ and\ \citenamefont
  {Roy}(2013)}]{henisch2013silicon}%
  \BibitemOpen
  \bibfield  {author} {\bibinfo {author} {\bibfnamefont {H.~K.}\ \bibnamefont
  {Henisch}}\ and\ \bibinfo {author} {\bibfnamefont {R.}~\bibnamefont {Roy}},\
  }\href@noop {} {\emph {\bibinfo {title} {Silicon Carbide 1968 - Proceedings
  of the International Conference on Silicon Carbide}}}\ (\bibinfo  {publisher}
  {Elsevier},\ \bibinfo {address} {New York},\ \bibinfo {year}
  {2013})\BibitemShut {NoStop}%
\bibitem [{\citenamefont {{Shur}}\ \emph {et~al.}(1999)\citenamefont {{Shur}},
  \citenamefont {{Levinshtein}},\ and\ \citenamefont
  {{Rumyantsev}}}]{Ioffe1999ShurEtAl-HandbookSeriesSemiconductorParametersVOL2}%
  \BibitemOpen
  \bibfield  {author} {\bibinfo {author} {\bibfnamefont {M.~S.}\ \bibnamefont
  {{Shur}}}, \bibinfo {author} {\bibfnamefont {M.}~\bibnamefont
  {{Levinshtein}}},\ and\ \bibinfo {author} {\bibfnamefont {S.}~\bibnamefont
  {{Rumyantsev}}},\ }\href {https://doi.org/10.1142/2046-vol2} {\emph {\bibinfo
  {title} {{Handbook Series on Semiconductor Parameters, Vol. 2: Ternary and
  Quaternary III-V Compounds}}}},\ Vol.~\bibinfo {volume} {2}\ (\bibinfo
  {publisher} {World Scientific Publishing Co},\ \bibinfo {address}
  {Singapore},\ \bibinfo {year} {1999})\ \bibinfo {note} {{See also:
  http://www.ioffe.ru/SVA/NSM/}}\BibitemShut {NoStop}%
\bibitem [{\citenamefont {Taylor}\ and\ \citenamefont
  {Yuan}(1962)}]{Taylor1962}%
  \BibitemOpen
  \bibfield  {author} {\bibinfo {author} {\bibfnamefont {T.}~\bibnamefont
  {Taylor}}\ and\ \bibinfo {author} {\bibfnamefont {F.}~\bibnamefont {Yuan}},\
  }\bibfield  {title} {\bibinfo {title} {Thermal stress and fracture in
  shear-constrained semiconductor device structures},\ }\href
  {https://doi.org/10.1109/t-ed.1962.14987} {\bibfield  {journal} {\bibinfo
  {journal} {{IRE} Transactions on Electron Devices}\ }\textbf {\bibinfo
  {volume} {9}},\ \bibinfo {pages} {303} (\bibinfo {year} {1962})}\BibitemShut
  {NoStop}%
\bibitem [{\citenamefont {Suhir}\ and\ \citenamefont
  {Vujosevic}(2008)}]{suhir2008interfacial}%
  \BibitemOpen
  \bibfield  {author} {\bibinfo {author} {\bibfnamefont {E.}~\bibnamefont
  {Suhir}}\ and\ \bibinfo {author} {\bibfnamefont {M.}~\bibnamefont
  {Vujosevic}},\ }\bibfield  {title} {\bibinfo {title} {Interfacial stresses in
  a bi-material assembly with a compliant bonding layer},\ }\href@noop {}
  {\bibfield  {journal} {\bibinfo  {journal} {Journal of Physics D: Applied
  Physics}\ }\textbf {\bibinfo {volume} {41}},\ \bibinfo {pages} {115504}
  (\bibinfo {year} {2008})}\BibitemShut {NoStop}%
\bibitem [{\citenamefont {B{\"u}ckle}\ \emph {et~al.}(2018)\citenamefont
  {B{\"u}ckle}, \citenamefont {Hauber}, \citenamefont {Cole}, \citenamefont
  {G{\"a}rtner}, \citenamefont {Zeimer}, \citenamefont {Grenzer},\ and\
  \citenamefont {Weig}}]{Bueckle2018APL-StressControl}%
  \BibitemOpen
  \bibfield  {author} {\bibinfo {author} {\bibfnamefont {M.}~\bibnamefont
  {B{\"u}ckle}}, \bibinfo {author} {\bibfnamefont {V.~C.}\ \bibnamefont
  {Hauber}}, \bibinfo {author} {\bibfnamefont {G.~D.}\ \bibnamefont {Cole}},
  \bibinfo {author} {\bibfnamefont {C.}~\bibnamefont {G{\"a}rtner}}, \bibinfo
  {author} {\bibfnamefont {U.}~\bibnamefont {Zeimer}}, \bibinfo {author}
  {\bibfnamefont {J.}~\bibnamefont {Grenzer}},\ and\ \bibinfo {author}
  {\bibfnamefont {E.~M.}\ \bibnamefont {Weig}},\ }\bibfield  {title} {\bibinfo
  {title} {Stress control of tensile-strained {In}$_{1-x}${Ga}$_{x}${P}
  nanomechanical string resonators},\ }\href
  {https://doi.org/10.1063/1.5054076} {\bibfield  {journal} {\bibinfo
  {journal} {Applied Physics Letters}\ }\textbf {\bibinfo {volume} {113}},\
  \bibinfo {pages} {201903} (\bibinfo {year} {2018})},\ \Eprint
  {https://arxiv.org/abs/https://doi.org/10.1063/1.5054076}
  {https://doi.org/10.1063/1.5054076} \BibitemShut {NoStop}%
\bibitem [{\citenamefont {Pakula}\ \emph {et~al.}(2004)\citenamefont {Pakula},
  \citenamefont {Yang}, \citenamefont {Pham}, \citenamefont {French},\ and\
  \citenamefont {Sarro}}]{pakula2004fabrication}%
  \BibitemOpen
  \bibfield  {author} {\bibinfo {author} {\bibfnamefont {L.}~\bibnamefont
  {Pakula}}, \bibinfo {author} {\bibfnamefont {H.}~\bibnamefont {Yang}},
  \bibinfo {author} {\bibfnamefont {H.}~\bibnamefont {Pham}}, \bibinfo {author}
  {\bibfnamefont {P.}~\bibnamefont {French}},\ and\ \bibinfo {author}
  {\bibfnamefont {P.}~\bibnamefont {Sarro}},\ }\bibfield  {title} {\bibinfo
  {title} {Fabrication of a cmos compatible pressure sensor for harsh
  environments},\ }\href {https://doi.org/10.1088/0960-1317/14/11/007}
  {\bibfield  {journal} {\bibinfo  {journal} {Journal of Micromechanics and
  Microengineering}\ }\textbf {\bibinfo {volume} {14}},\ \bibinfo {pages}
  {1478} (\bibinfo {year} {2004})}\BibitemShut {NoStop}%
\bibitem [{\citenamefont {Zorman}\ and\ \citenamefont
  {Parro}(2008)}]{zorman2008micro}%
  \BibitemOpen
  \bibfield  {author} {\bibinfo {author} {\bibfnamefont {C.~A.}\ \bibnamefont
  {Zorman}}\ and\ \bibinfo {author} {\bibfnamefont {R.~J.}\ \bibnamefont
  {Parro}},\ }\bibfield  {title} {\bibinfo {title} {Micro-and nanomechanical
  structures for silicon carbide mems and nems},\ }\href
  {https://doi.org/https://doi.org/10.1002/pssb.200844135} {\bibfield
  {journal} {\bibinfo  {journal} {physica status solidi (b)}\ }\textbf
  {\bibinfo {volume} {245}},\ \bibinfo {pages} {1404} (\bibinfo {year}
  {2008})}\BibitemShut {NoStop}%
\bibitem [{\citenamefont {Iacopi}\ \emph {et~al.}(2013)\citenamefont {Iacopi},
  \citenamefont {Brock}, \citenamefont {Iacopi}, \citenamefont {Hold},\ and\
  \citenamefont {Dauskardt}}]{iacopi2013evidence}%
  \BibitemOpen
  \bibfield  {author} {\bibinfo {author} {\bibfnamefont {F.}~\bibnamefont
  {Iacopi}}, \bibinfo {author} {\bibfnamefont {R.~E.}\ \bibnamefont {Brock}},
  \bibinfo {author} {\bibfnamefont {A.}~\bibnamefont {Iacopi}}, \bibinfo
  {author} {\bibfnamefont {L.}~\bibnamefont {Hold}},\ and\ \bibinfo {author}
  {\bibfnamefont {R.~H.}\ \bibnamefont {Dauskardt}},\ }\bibfield  {title}
  {\bibinfo {title} {Evidence of a highly compressed nanolayer at the epitaxial
  silicon carbide interface with silicon},\ }\href
  {https://doi.org/https://doi.org/10.1016/j.actamat.2013.07.034} {\bibfield
  {journal} {\bibinfo  {journal} {Acta Materialia}\ }\textbf {\bibinfo {volume}
  {61}},\ \bibinfo {pages} {6533} (\bibinfo {year} {2013})}\BibitemShut
  {NoStop}%
\bibitem [{\citenamefont {{Iacopi}}\ \emph {et~al.}(2013)\citenamefont
  {{Iacopi}}, \citenamefont {{Walker}}, \citenamefont {{Wang}}, \citenamefont
  {{Malesys}}, \citenamefont {{Ma}}, \citenamefont {{Cunning}},\ and\
  \citenamefont
  {{Iacopi}}}]{Iacopi2013APL-QrientationDependentStressRelax3C-SiCFilms}%
  \BibitemOpen
  \bibfield  {author} {\bibinfo {author} {\bibfnamefont {F.}~\bibnamefont
  {{Iacopi}}}, \bibinfo {author} {\bibfnamefont {G.}~\bibnamefont {{Walker}}},
  \bibinfo {author} {\bibfnamefont {L.}~\bibnamefont {{Wang}}}, \bibinfo
  {author} {\bibfnamefont {L.}~\bibnamefont {{Malesys}}}, \bibinfo {author}
  {\bibfnamefont {S.}~\bibnamefont {{Ma}}}, \bibinfo {author} {\bibfnamefont
  {B.~V.}\ \bibnamefont {{Cunning}}},\ and\ \bibinfo {author} {\bibfnamefont
  {A.}~\bibnamefont {{Iacopi}}},\ }\bibfield  {title} {\bibinfo {title}
  {{Orientation-dependent stress relaxation in hetero-epitaxial 3{C}-{SiC}
  films}},\ }\href {https://doi.org/10.1063/1.4774087} {\bibfield  {journal}
  {\bibinfo  {journal} {Applied Physics Letters}\ }\textbf {\bibinfo {volume}
  {102}},\ \bibinfo {eid} {011908} (\bibinfo {year} {2013})}\BibitemShut
  {NoStop}%
\bibitem [{\citenamefont {{Kermany}}\ \emph {et~al.}(2016)\citenamefont
  {{Kermany}}, \citenamefont {{Bennett}}, \citenamefont {{Brawley}},
  \citenamefont {{Bowen}},\ and\ \citenamefont
  {{Iacopi}}}]{Kermany2016JAP-FactorsAffectionFQproduct3CSiC}%
  \BibitemOpen
  \bibfield  {author} {\bibinfo {author} {\bibfnamefont {A.~R.}\ \bibnamefont
  {{Kermany}}}, \bibinfo {author} {\bibfnamefont {J.~S.}\ \bibnamefont
  {{Bennett}}}, \bibinfo {author} {\bibfnamefont {G.~A.}\ \bibnamefont
  {{Brawley}}}, \bibinfo {author} {\bibfnamefont {W.~P.}\ \bibnamefont
  {{Bowen}}},\ and\ \bibinfo {author} {\bibfnamefont {F.}~\bibnamefont
  {{Iacopi}}},\ }\bibfield  {title} {\bibinfo {title} {Factors affecting the f
  {$\times$} {Q} product of 3{C}-{SiC} microstrings: What is the upper limit
  for sensitivity?},\ }\href {https://doi.org/10.1063/1.4941274} {\bibfield
  {journal} {\bibinfo  {journal} {Journal of Applied Physics}\ }\textbf
  {\bibinfo {volume} {119}},\ \bibinfo {eid} {055304} (\bibinfo {year}
  {2016})}\BibitemShut {NoStop}%
\bibitem [{\citenamefont {Romero}\ \emph {et~al.}(2020)\citenamefont {Romero},
  \citenamefont {Valenzuela}, \citenamefont {Kermany}, \citenamefont
  {Sementilli}, \citenamefont {Iacopi},\ and\ \citenamefont
  {Bowen}}]{Romero2020}%
  \BibitemOpen
  \bibfield  {author} {\bibinfo {author} {\bibfnamefont {E.}~\bibnamefont
  {Romero}}, \bibinfo {author} {\bibfnamefont {V.~M.}\ \bibnamefont
  {Valenzuela}}, \bibinfo {author} {\bibfnamefont {A.~R.}\ \bibnamefont
  {Kermany}}, \bibinfo {author} {\bibfnamefont {L.}~\bibnamefont {Sementilli}},
  \bibinfo {author} {\bibfnamefont {F.}~\bibnamefont {Iacopi}},\ and\ \bibinfo
  {author} {\bibfnamefont {W.~P.}\ \bibnamefont {Bowen}},\ }\bibfield  {title}
  {\bibinfo {title} {Engineering the dissipation of crystalline micromechanical
  resonators},\ }\href@noop {} {\bibfield  {journal} {\bibinfo  {journal}
  {Physical Review Applied}\ }\textbf {\bibinfo {volume} {13}},\ \bibinfo
  {pages} {044007} (\bibinfo {year} {2020})}\BibitemShut {NoStop}%
\bibitem [{\citenamefont {Bereyhi}\ \emph {et~al.}(2019)\citenamefont
  {Bereyhi}, \citenamefont {Beccari}, \citenamefont {Fedorov}, \citenamefont
  {Ghadimi}, \citenamefont {Schilling}, \citenamefont {Wilson}, \citenamefont
  {Engelsen},\ and\ \citenamefont {Kippenberg}}]{Bereyhi2019}%
  \BibitemOpen
  \bibfield  {author} {\bibinfo {author} {\bibfnamefont {M.~J.}\ \bibnamefont
  {Bereyhi}}, \bibinfo {author} {\bibfnamefont {A.}~\bibnamefont {Beccari}},
  \bibinfo {author} {\bibfnamefont {S.~A.}\ \bibnamefont {Fedorov}}, \bibinfo
  {author} {\bibfnamefont {A.~H.}\ \bibnamefont {Ghadimi}}, \bibinfo {author}
  {\bibfnamefont {R.}~\bibnamefont {Schilling}}, \bibinfo {author}
  {\bibfnamefont {D.~J.}\ \bibnamefont {Wilson}}, \bibinfo {author}
  {\bibfnamefont {N.~J.}\ \bibnamefont {Engelsen}},\ and\ \bibinfo {author}
  {\bibfnamefont {T.~J.}\ \bibnamefont {Kippenberg}},\ }\bibfield  {title}
  {\bibinfo {title} {Clamp-tapering increases the quality factor of stressed
  nanobeams},\ }\href {https://doi.org/10.1021/acs.nanolett.8b04942} {\bibfield
   {journal} {\bibinfo  {journal} {Nano Letters}\ }\textbf {\bibinfo {volume}
  {19}},\ \bibinfo {pages} {2329} (\bibinfo {year} {2019})}\BibitemShut
  {NoStop}%
\bibitem [{\citenamefont {{Unterreithmeier}}\ \emph {et~al.}(2009)\citenamefont
  {{Unterreithmeier}}, \citenamefont {{Weig}},\ and\ \citenamefont
  {{Kotthaus}}}]{Unterrreithmeier2009Nature-TransductionDielectricForces}%
  \BibitemOpen
  \bibfield  {author} {\bibinfo {author} {\bibfnamefont {Q.~P.}\ \bibnamefont
  {{Unterreithmeier}}}, \bibinfo {author} {\bibfnamefont {E.~M.}\ \bibnamefont
  {{Weig}}},\ and\ \bibinfo {author} {\bibfnamefont {J.~P.}\ \bibnamefont
  {{Kotthaus}}},\ }\bibfield  {title} {\bibinfo {title} {{Universal
  transduction scheme for nanomechanical systems based ob dielectric forces}},\
  }\href@noop {} {\bibfield  {journal} {\bibinfo  {journal} {Nature}\ }\textbf
  {\bibinfo {volume} {458}} (\bibinfo {year} {2009})}\BibitemShut {NoStop}%
\bibitem [{\citenamefont {Yang}\ and\ \citenamefont {Paul}(2002)}]{Yang2002}%
  \BibitemOpen
  \bibfield  {author} {\bibinfo {author} {\bibfnamefont {J.}~\bibnamefont
  {Yang}}\ and\ \bibinfo {author} {\bibfnamefont {O.}~\bibnamefont {Paul}},\
  }\bibfield  {title} {\bibinfo {title} {Fracture properties of {LPCVD} silicon
  nitride thin films from the load{\textendash}deflection of long membranes},\
  }\href {https://doi.org/10.1016/s0924-4247(02)00049-3} {\bibfield  {journal}
  {\bibinfo  {journal} {Sensors and Actuators A: Physical}\ }\textbf {\bibinfo
  {volume} {97-98}},\ \bibinfo {pages} {520} (\bibinfo {year}
  {2002})}\BibitemShut {NoStop}%
\bibitem [{\citenamefont {Kaushik}\ \emph {et~al.}(2005)\citenamefont
  {Kaushik}, \citenamefont {Kahn},\ and\ \citenamefont {Heuer}}]{Kaushik2005}%
  \BibitemOpen
  \bibfield  {author} {\bibinfo {author} {\bibfnamefont {A.}~\bibnamefont
  {Kaushik}}, \bibinfo {author} {\bibfnamefont {H.}~\bibnamefont {Kahn}},\ and\
  \bibinfo {author} {\bibfnamefont {A.}~\bibnamefont {Heuer}},\ }\bibfield
  {title} {\bibinfo {title} {Wafer-level mechanical characterization of silicon
  nitride {MEMS}},\ }\href {https://doi.org/10.1109/jmems.2004.839315}
  {\bibfield  {journal} {\bibinfo  {journal} {Journal of Microelectromechanical
  Systems}\ }\textbf {\bibinfo {volume} {14}},\ \bibinfo {pages} {359}
  (\bibinfo {year} {2005})}\BibitemShut {NoStop}%
\bibitem [{\citenamefont {Petersen}(1982)}]{Petersen1982}%
  \BibitemOpen
  \bibfield  {author} {\bibinfo {author} {\bibfnamefont {K.}~\bibnamefont
  {Petersen}},\ }\bibfield  {title} {\bibinfo {title} {Silicon as a mechanical
  material},\ }\href {https://doi.org/10.1109/proc.1982.12331} {\bibfield
  {journal} {\bibinfo  {journal} {Proceedings of the {IEEE}}\ }\textbf
  {\bibinfo {volume} {70}},\ \bibinfo {pages} {420} (\bibinfo {year}
  {1982})}\BibitemShut {NoStop}%
\bibitem [{zen()}]{zenodo}%
  \BibitemOpen
  \href@noop {} {\bibinfo {title} {Data and analysis code are available at
  \url{https://doi.org/10.5281/zenodo.4588507}.}}\BibitemShut {Stop}%
\end{thebibliography}%


\begin{thebibliography}{6}%
\makeatletter
\providecommand \@ifxundefined [1]{%
 \@ifx{#1\undefined}
}%
\providecommand \@ifnum [1]{%
 \ifnum #1\expandafter \@firstoftwo
 \else \expandafter \@secondoftwo
 \fi
}%
\providecommand \@ifx [1]{%
 \ifx #1\expandafter \@firstoftwo
 \else \expandafter \@secondoftwo
 \fi
}%
\providecommand \natexlab [1]{#1}%
\providecommand \enquote  [1]{``#1''}%
\providecommand \bibnamefont  [1]{#1}%
\providecommand \bibfnamefont [1]{#1}%
\providecommand \citenamefont [1]{#1}%
\providecommand \href@noop [0]{\@secondoftwo}%
\providecommand \href [0]{\begingroup \@sanitize@url \@href}%
\providecommand \@href[1]{\@@startlink{#1}\@@href}%
\providecommand \@@href[1]{\endgroup#1\@@endlink}%
\providecommand \@sanitize@url [0]{\catcode `\\12\catcode `\$12\catcode
  `\&12\catcode `\#12\catcode `\^12\catcode `\_12\catcode `\%12\relax}%
\providecommand \@@startlink[1]{}%
\providecommand \@@endlink[0]{}%
\providecommand \url  [0]{\begingroup\@sanitize@url \@url }%
\providecommand \@url [1]{\endgroup\@href {#1}{\urlprefix }}%
\providecommand \urlprefix  [0]{URL }%
\providecommand \Eprint [0]{\href }%
\providecommand \doibase [0]{https://doi.org/}%
\providecommand \selectlanguage [0]{\@gobble}%
\providecommand \bibinfo  [0]{\@secondoftwo}%
\providecommand \bibfield  [0]{\@secondoftwo}%
\providecommand \translation [1]{[#1]}%
\providecommand \BibitemOpen [0]{}%
\providecommand \bibitemStop [0]{}%
\providecommand \bibitemNoStop [0]{.\EOS\space}%
\providecommand \EOS [0]{\spacefactor3000\relax}%
\providecommand \BibitemShut  [1]{\csname bibitem#1\endcsname}%
\let\auto@bib@innerbib\@empty
\bibitem [{\citenamefont {Doster}()}]{DosterUnpub}%
  \BibitemOpen
  \bibfield  {author} {\bibinfo {author} {\bibfnamefont {J.}~\bibnamefont
  {Doster}},\ }\href@noop {} {\bibinfo  {journal} {unpublished}\ }\BibitemShut
  {NoStop}%
\bibitem [{\citenamefont {Maluf}\ and\ \citenamefont
  {Williams}(2004)}]{maluf2002introduction}%
  \BibitemOpen
\bibfield  {journal} {  }\bibfield  {author} {\bibinfo {author} {\bibfnamefont
  {N.}~\bibnamefont {Maluf}}\ and\ \bibinfo {author} {\bibfnamefont
  {K.}~\bibnamefont {Williams}},\ }\href@noop {} {\emph {\bibinfo {title} {An
  introduction to microelectromechanical systems engineering}}}\ (\bibinfo
  {publisher} {Artech House},\ \bibinfo {address} {Norwood},\ \bibinfo {year}
  {2004})\BibitemShut {NoStop}%
\bibitem [{\citenamefont {Li}\ and\ \citenamefont
  {Bradt}(1987)}]{li1987single}%
  \BibitemOpen
  \bibfield  {author} {\bibinfo {author} {\bibfnamefont {Z.}~\bibnamefont
  {Li}}\ and\ \bibinfo {author} {\bibfnamefont {R.~C.}\ \bibnamefont {Bradt}},\
  }\bibfield  {title} {\bibinfo {title} {The single-crystal elastic constants
  of cubic (3c) sic to 1000 c},\ }\href@noop {} {\bibfield  {journal} {\bibinfo
   {journal} {Journal of materials science}\ }\textbf {\bibinfo {volume}
  {22}},\ \bibinfo {pages} {2557} (\bibinfo {year} {1987})}\BibitemShut
  {NoStop}%
\bibitem [{\citenamefont {Henisch}\ and\ \citenamefont
  {Roy}(2013)}]{henisch2013silicon}%
  \BibitemOpen
  \bibfield  {author} {\bibinfo {author} {\bibfnamefont {H.~K.}\ \bibnamefont
  {Henisch}}\ and\ \bibinfo {author} {\bibfnamefont {R.}~\bibnamefont {Roy}},\
  }\href@noop {} {\emph {\bibinfo {title} {Silicon Carbide 1968 - Proceedings
  of the International Conference on Silicon Carbide}}}\ (\bibinfo  {publisher}
  {Elsevier},\ \bibinfo {address} {New York},\ \bibinfo {year}
  {2013})\BibitemShut {NoStop}%
\bibitem [{\citenamefont {{Shur}}\ \emph {et~al.}(1999)\citenamefont {{Shur}},
  \citenamefont {{Levinshtein}},\ and\ \citenamefont
  {{Rumyantsev}}}]{Ioffe1999ShurEtAl-HandbookSeriesSemiconductorParametersVOL2}%
  \BibitemOpen
  \bibfield  {author} {\bibinfo {author} {\bibfnamefont {M.~S.}\ \bibnamefont
  {{Shur}}}, \bibinfo {author} {\bibfnamefont {M.}~\bibnamefont
  {{Levinshtein}}},\ and\ \bibinfo {author} {\bibfnamefont {S.}~\bibnamefont
  {{Rumyantsev}}},\ }\href {https://doi.org/10.1142/2046-vol2} {\emph {\bibinfo
  {title} {{Handbook Series on Semiconductor Parameters, Vol. 2: Ternary and
  Quaternary III-V Compounds}}}},\ Vol.~\bibinfo {volume} {2}\ (\bibinfo
  {publisher} {World Scientific Publishing Co},\ \bibinfo {address}
  {Singapore},\ \bibinfo {year} {1999})\ \bibinfo {note} {{See also:
  http://www.ioffe.ru/SVA/NSM/}}\BibitemShut {NoStop}%
\bibitem [{\citenamefont {B{\"u}ckle}\ \emph {et~al.}(2018)\citenamefont
  {B{\"u}ckle}, \citenamefont {Hauber}, \citenamefont {Cole}, \citenamefont
  {G{\"a}rtner}, \citenamefont {Zeimer}, \citenamefont {Grenzer},\ and\
  \citenamefont {Weig}}]{Bueckle2018APL-StressControl}%
  \BibitemOpen
  \bibfield  {author} {\bibinfo {author} {\bibfnamefont {M.}~\bibnamefont
  {B{\"u}ckle}}, \bibinfo {author} {\bibfnamefont {V.~C.}\ \bibnamefont
  {Hauber}}, \bibinfo {author} {\bibfnamefont {G.~D.}\ \bibnamefont {Cole}},
  \bibinfo {author} {\bibfnamefont {C.}~\bibnamefont {G{\"a}rtner}}, \bibinfo
  {author} {\bibfnamefont {U.}~\bibnamefont {Zeimer}}, \bibinfo {author}
  {\bibfnamefont {J.}~\bibnamefont {Grenzer}},\ and\ \bibinfo {author}
  {\bibfnamefont {E.~M.}\ \bibnamefont {Weig}},\ }\bibfield  {title} {\bibinfo
  {title} {Stress control of tensile-strained {In}$_{1-x}${Ga}$_{x}${P}
  nanomechanical string resonators},\ }\href
  {https://doi.org/10.1063/1.5054076} {\bibfield  {journal} {\bibinfo
  {journal} {Applied Physics Letters}\ }\textbf {\bibinfo {volume} {113}},\
  \bibinfo {pages} {201903} (\bibinfo {year} {2018})},\ \Eprint
  {https://arxiv.org/abs/https://doi.org/10.1063/1.5054076}
  {https://doi.org/10.1063/1.5054076} \BibitemShut {NoStop}%
\end{thebibliography}%

\end{document}


\title{Supplemental Material: Universal Length Dependence of Tensile Stress in Nanomechanical String Resonators}
	
	\author{Maximilian B{\"u}ckle}
	\altaffiliation{Both authors contributed equally to this work.}
	\affiliation{%
		Department of Physics, University of Konstanz, D-78457 Konstanz, Germany
	}%
	
	\author{Yannick S. Kla{\ss}}%
	\altaffiliation{Both authors contributed equally to this work.}
	\affiliation{%
		Department of Physics, University of Konstanz, D-78457 Konstanz, Germany
	}%
	
	\author{Felix B. N{\"a}gele}%
	\affiliation{%
		Department of Physics, University of Konstanz, D-78457 Konstanz, Germany
	}%
	
	\author{R\'{e}my Braive}
	\affiliation{%
		Centre de Nanosciences et de Nanotechnologies, CNRS, Universit\'{e} Paris-Sud, Universit\'{e} Paris-Saclay, 91767 Palaiseau, France
	}%
	
	\author{Eva M. Weig}
	\email{eva.weig@uni-konstanz.de, eva.weig@tum.de}
	\affiliation{%
		Department of Physics, University of Konstanz, D-78457 Konstanz, Germany
	}%
	\affiliation{%
		Department of Electrical and Computer Engineering, Technical University Munich, D-80939 Munich, Germany
	}%

	\maketitle

	\appendix

	This Supplemental Material provides further information about wafer and material parameters in Appendix~\ref{A}, the finite element simulations performed to validate the geometric reconfiguration of a fully released nanostring in Appendix~\ref{B}, and the experimental confirmation of the shearing contraction of the clamping pedestals in Appendix~\ref{C}.

	\section{Wafers and material parameters}	
	\label{A}
	
	In Tab. \ref{tab:wafers} the growth parameters of the four wafers employed in this work are summarized, stating the thickness of the device layer, sacrificial layer (if the system has one), substrate, and the corresponding supplier. The two SiN wafers were grown by Low Pressure Chemical Vapor Deposition (LPCVD), the SiC in a two-stage Chemical Vapor Deposition process, and the InGaP using Metal-Organic Chemical Vapor Deposition (MOCVD).

	\begin{table*}[h!]
		\caption{
			\label{tab:wafers}
			Basic parameters of the wafers on which the string resonators were fabricated.
		}
		\begin{ruledtabular}
			\begin{tabular}{lllll}
				& resonator/device layer & sacrificial layer & substrate & source \\
				\hline
				SiN-FS 			& 100\,nm SiN	& --- 	& SiO$_2$ & HSG-IMIT  \\
				SiN-Si 			& 100\,nm SiN 	& 400\,nm SiO$_2$ 	& Si(100) & HSG-IMIT  \\
				SiC 			& 110\,nm 3C-SiC 	& --- 	& Si(111) & NOVASiC  \\
				InGaP 	& \multicolumn{1}{r}{100\,nm In$_{0.415}$Ga$_{0.585}$P} 	& \multicolumn{1}{l}{1000\,nm Al$_{0.85}$Ga$_{0.15}$As} 	& GaAs & CNRS  \\
			\end{tabular}
		\end{ruledtabular}
	\end{table*}
	
	\newpage
	
	All material parameters employed in the theoretical calculations, i.e. the Young's modulus $ E $, the shear modulus $ G $, the Poisson ratio $ \nu $, and the mass density $\rho$ of the respective materials, are listed in Tab. \ref{tab:ModelParameters}. 
	
	\begin{table}[h]
		\caption{
			\label{tab:ModelParameters}
			Young's modulus, shear modulus and density of the materials used within the main text. All shear mouduli where calculated via $ G = \frac{E}{2(1+\nu)} $ . 
		}
		\begin{ruledtabular}
			\begin{tabular}{crrrr}
				& \multicolumn{1}{c}{Young's modulus $ E $} & \multicolumn{1}{c}{Shear modulus $ G $}& \multicolumn{1}{c}{Poisson's ratio $ \nu $} &\multicolumn{1}{c}{density $\rho$} \\
				& \multicolumn{1}{c}{\textrm{(GPa)}} & \multicolumn{1}{c}{\textrm{(GPa)}} & & \multicolumn{1}{c}{(g/cm$ ^3 $)} \\
				\hline
				SiN	& 260 \cite{DosterUnpub} & 104 & 0.25\cite{maluf2002introduction} & 3.1 \cite{maluf2002introduction}	  \\
				SiO$_2$	& 73 \cite{maluf2002introduction} & 31 & 0.17 \cite{maluf2002introduction} &2.2 \cite{maluf2002introduction}	  \\
				Si & 160 \cite{maluf2002introduction} & 66 & 0.22 \cite{maluf2002introduction} &2.4 \cite{maluf2002introduction}	  \\
				SiC	& 419 \cite{li1987single}\footnotemark[1] &	184 & 0.14 \cite{maluf2002introduction} &3.166 \cite{henisch2013silicon}\\ 
				InGaP	& 124 \cite{Ioffe1999ShurEtAl-HandbookSeriesSemiconductorParametersVOL2}\footnotemark[1]  & 47 & 0.32\cite{Ioffe1999ShurEtAl-HandbookSeriesSemiconductorParametersVOL2}\footnotemark[2] & 4.418 \cite{Ioffe1999ShurEtAl-HandbookSeriesSemiconductorParametersVOL2}	  \\
				GaAs	& 75  \cite{maluf2002introduction} & 29 & 0.31 \cite{Ioffe1999ShurEtAl-HandbookSeriesSemiconductorParametersVOL2} &5.3 \cite{maluf2002introduction}	  \\
			\end{tabular}
		\end{ruledtabular}
		\footnotemark[1]{Determined by utilizing the elastic constants and the stiffness tensor as described in \cite{Bueckle2018APL-StressControl}.}
		\footnotemark[2]{Calculated with $ \nu = \frac{c_{12}}{c_{11}+c_{12}} $ where $ c_{ij} $ are the elastic constants.}
		\vspace{-4mm}
	\end{table}

	\section{Finite Element Method Simulations}
	\label{B}
	
\begin{figure}[h!]
		\includegraphics[width=0.99\linewidth]{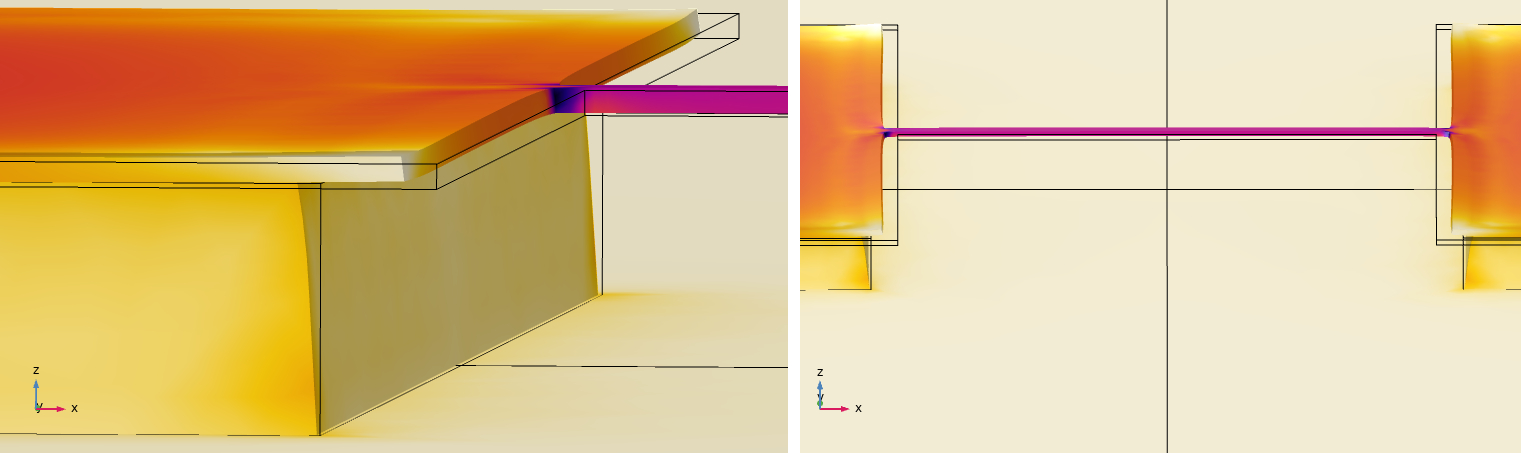}
		\caption{ FEM simulations of a single string resonator with an initial stress of $2.9$\,GPa. Furthermore, we set a thickness of $100$\,nm for the device layer, a $500$\,nm undercut and a pedestal height of $1\,\mu $m.
			\label{fig:femSimulationPedestal} 
		}
	\end{figure}

	The geometric reconfiguration of the pedestal, the clamping pad and the string was explored in more detail by finite element method (FEM) simulations to validate our theoretical considerations. To this end, the individual $10\,\mu$m long and $300$\,nm wide SiN-FS string resonator held in place by two SiO$ _2 $ pedestals on a SiO$_2$ substrate shown in Fig.~\ref{fig:femSimulationPedestal} was considered. The thickness of the device layer was set to $100$\,nm, a $500$\,nm undercut and a pedestal height of $1\,\mu$m were assumed, as well as an initial two-dimensional tensile stress of $2.9$\,GPa. A perfectly matched layer was included to mimick an infinite substrate, but did not noticeably influence the result.
	A close look at Fig.~\ref{fig:femSimulationPedestal} clearly reveals the shearing of the pedestal as well as the contraction of the clamping pad due to the stressed device layer. Also apparent is the resulting elongation of the string and the enhanced tensile stress in the string, which, for the case of the extremely short length of the simulated string, even exceeds the remaining tensile stress in the clamping pad. 
	These observations qualitatively support all assumptions of the elastic model put forward in the main text.\\
	
	\section{Measuring the pedestal contraction}
	\label{C}
	
\begin{figure}[h!]
		\includegraphics[width=0.9\linewidth]{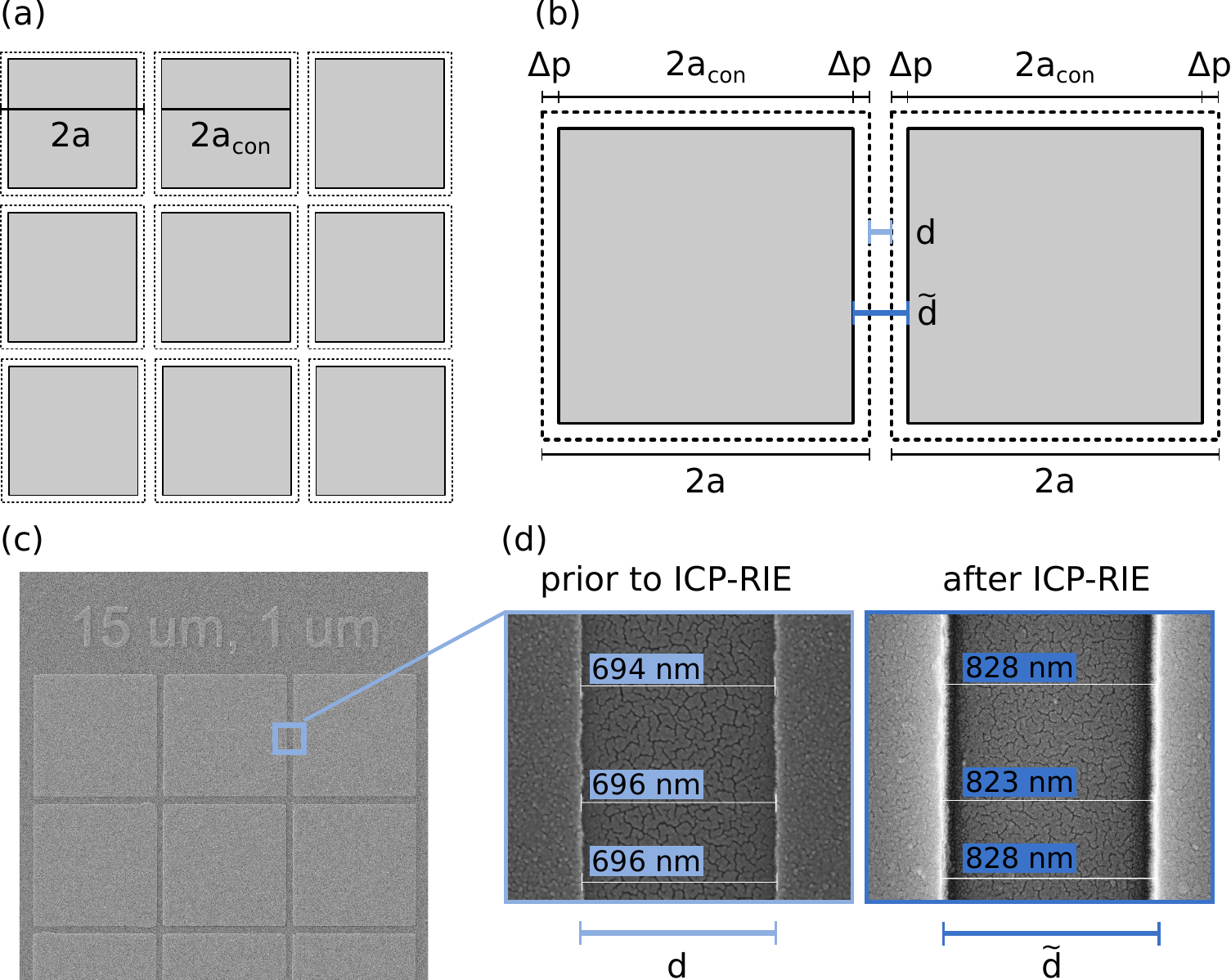}
		\caption{ Array of pedestals (a) and a close up (b) including length annotations. Dashed lines and solid lines correspond to the pedestal before and after contraction, respectively. (c) SEM image of the array structure before it was etched. (d) SEM image of the gap between two pedestals before (left) and after (right) contraction.
			\label{fig:MeasurePedCont} 
		}
	\end{figure}

	To further support our elastic model, we have experimentally quantified the shearing of the pedestal using the test structures discussed in the following.
	An array of quadratic pedestals is fabricated on SiN-FS (see Fig.~\ref{fig:MeasurePedCont}(a) and (c)), the material for which the biggest contraction is expected (c.f. Tab. \RNum{2}). As shown in Fig.~\ref{fig:MeasurePedCont}(a) and (b), the uncontracted width of a pedestal is $ 2 a $ and the pedestal-pedestal distance is $ d $. An anistropic ICP-RIE etch step (etching depth of around 350~nm) allows for a contraction of the pedestal by $ 2 \Delta p $ to $ 2 a_{\mathrm{con}} = 2 a - 2 \Delta p$. Because the contraction is in the nanometer regime and the pedestal in the micrometer regime, we can not simply image the whole pedestal and directly measure $ 2a $ and $ 2a_{\mathrm{con}}  $ and calculate the contraction $ 2 \Delta p $, as this is beyond the resolution of our scanning electron microscopy. However, as indicated schematically in Fig. ~\ref{fig:MeasurePedCont}(b), the separation of two closely-spaced pedestals of the test structure can be mapped out with a higher resolution. Comparison of their spacing before and after the contraction, $ d $ and $ \tilde{d} $, respectively, indeed yields in increase of the gap as shown in Fig. ~\ref{fig:MeasurePedCont}(d), which indicates a contraction of the clamping structure. For our sample chip we measure an average value of $ d = 793(6) $~nm and $ \tilde{d} = 824 (7) $~nm (over the entire gap, not just the section close to the center shown in Fig. ~\ref{fig:MeasurePedCont}(d)), corresponding to a contraction of  $ \Delta p = \frac{\tilde{d}-d}{2} = 15(7)~ $nm. The theoretical value of $ \Delta p = 8~ $nm is just within the uncertainty of the measured value.

	\clearpage
	
	\bibliography{si}